\documentclass[runningheads]{llncs}
\usepackage[T1]{fontenc}
\usepackage{graphicx}
\usepackage{tikz}
\usepackage{lipsum}
\usepackage{subcaption}
\usetikzlibrary{decorations.pathreplacing, arrows, fit}
\usetikzlibrary{arrows.meta}
\begin{document}

\makeatletter
\renewcommand\section{\@startsection{section}{1}{\z@}%
                       {-10\p@ \@plus -4\p@ \@minus -4\p@}%
                       {6\p@ \@plus 4\p@ \@minus 4\p@}%
                       {\normalfont\large\bfseries\boldmath
                        \rightskip=\z@ \@plus 8em\pretolerance=10000 }}
\renewcommand\subsection{\@startsection{subsection}{2}{\z@}%
                       {-10\p@ \@plus -4\p@ \@minus -4\p@}%
                       {6\p@ \@plus 4\p@ \@minus 4\p@}%
                       {\normalfont\normalsize\bfseries\boldmath
                        \rightskip=\z@ \@plus 8em\pretolerance=10000 }}
\renewcommand\subsubsection{\@startsection{subsubsection}{3}{\z@}%
                       {-6\p@ \@plus -4\p@ \@minus -4\p@}%
                       {-1.5em \@plus -0.22em \@minus -0.1em}%
                       {\normalfont\normalsize\bfseries\boldmath}}
\makeatother
\interlinepenalty=-1
\setlength{\textfloatsep}{5pt}

\setlength{\abovedisplayskip}{3pt}
\setlength{\belowdisplayskip}{3pt}
%\title{Ethereum's Limited Transaction Parallelizability}
\title{DeFi and NFTs Hinder Blockchain Scalability}
% Why current NFTs and DeFi thwart/hinder Blockchain/Ethereum Scaling/Scalability

%\title{Illustrating Ethereum's Limited  Parallelizability}
%Ethereum's Unacceptable State of Parallelization Potential
%Why Ethereum's Parallelization Potential is Limited?
%Ethereum's Limited Parallelization Potential
\author{Lioba Heimbach, Quentin Kniep, Yann Vonlanthen, Roger Wattenhofer}
\institute{ETH Zurich\\Zurich, Switzerland}
\authorrunning{Heimbach et al.}

%
\maketitle              % typeset the header of the contribution
\begin{abstract}

Many classical blockchains are known to have an embarrassingly low transaction throughput, down to Bitcoin's notorious seven transactions per second limit. Various proposals and implementations for increasing throughput emerged in the first decade of blockchain research. But how much concurrency is possible? In their early days, blockchains were mostly used for simple transfers from user to user. More recently, however, \emph{decentralized finance (DeFi)} and \emph{NFT marketplaces} have completely changed what is happening on blockchains. Both are built using \emph{smart contracts} and have gained significant popularity. Transactions on DeFi and NFT marketplaces often interact with the same smart contracts. We believe this development has transformed blockchain usage.

In our work, we perform a historical analysis of Ethereum's transaction graph. We study how much interaction between transactions there was historically and how much there is now. We find that the rise of DeFi and NFT marketplaces has led to an increase in ``centralization'' in the transaction graph. More transactions are now interconnected: currently there are around 200 transactions per block with 4000 interdependencies between them. We further find that the parallelizability of Ethereum's current interconnected transaction workload is limited. A speedup exceeding a factor of five is currently unrealistic. 

%In fact, we identify that some of DeFi's core smart contracts are central to the increased connectedness of Ethereum's transaction graph and, even more astonishing, introduce unnecessary dependencies. This limits the potential for concurrency and parallelizability of transactions. To tackle these avoidable dependencies, we introduce and perform a disentanglement of the transaction data. However, we find that Ethereum's parallelizability is limited even for the disentangled transaction data. Our work concludes that it is not sufficient to propose scaling solutions, but that some efforts should instead be shifted to reducing the number of transaction dependencies in the first place. 

%TODO go over keywords
\end{abstract}
\section{Introduction}

When the first blockchain, Bitcoin~\cite{nakamoto2008bitcoin}, was launched in 2008, it allowed the execution of financial transactions without relying on a central authority. With its promise, Bitcoin sparked the creation of many more cryptocurrencies, most notably Ethereum~\cite{wood2014ethereum}, which introduced smart contracts in 2015. However, even though cryptocurrencies are continuously reaching new levels of popularity, the transaction throughput of the most popular\footnote{Measured by total fees users are willing to pay to use the blockchain (see https://cryptofees.info) Ethereum is orders of magnitude more popular than other smart contract-enabled blockchains, such as Avalanche and Cardano.} ones remains incredibly low.

Given the low throughput of blockchains, especially in comparison to established payment systems such as Visa or PayPal, many suggestions to tackle low blockchain throughput levels have been introduced as well as implemented. Layer 2 protocols~\cite{Decker2015fast,poon2016bitcoin,kalodner2018arbitrum,2022optimism}, handling transactions off-chain, sharding protocols~\cite{2022sharding,han2020analysing,luu2016secure,zamani2018rapidchain,dang2019towards,wang2019sok,avarikioti2019divide}, and moving from \emph{Proof-of-Work (PoW)} to \emph{Proof-of-Stake (PoS)}~\cite{2022merge} are amongst the most adopted scaling solutions. In addition to the development of the aforementioned solutions, the potential of concurrency control for multithreaded execution has been explored thoroughly.

However, these solutions do not focus on the implications of the changing nature of blockchain transactions. Before the rapid rise of \emph{decentralized finance (DeFi)} and NFT marketplaces, transactions were largely simple transactions between two parties. Consequently, dependencies between a large set of transactions in a block were rare. In the face of few dependencies, transaction throughput can be increased with the proposed solutions -- as they rely on the parallel execution of transactions to increase throughput. However, DeFi and NFT marketplaces have brought new challenges when scaling throughput.

While DeFi employs smart contracts hosted on the blockchain to offer many of the services provided by traditional finance, NFT marketplaces utilize smart contracts to facilitate NFT purchases. Core smart contracts building DeFi and NFT marketplaces are involved in many of a block's transactions and create dependencies between a significant proportion of transactions. This new reality on Ethereum greatly challenges the parallelization of transaction execution. 

In this work, we explore the limits of transaction parallelization on Ethereum. %, which is by far the largest blockchain by transaction volume, when also considering smart contract activity. 
We analyze these limits by investigating the connectedness of the Ethereum mainnet transaction graph over time. The identification of the largest connected component and clique in terms of the required execution workload in a block's transaction graph allows us to explore the potential of concurrent execution over time. 
In particular, we point out that DeFi's most important smart contracts are central in the transaction graph and responsible for the vast majority of transaction dependencies. Thus, a handful of smart contracts present a significant parallelization bottleneck, especially given the widespread adoption of DeFi and NFT marketplaces starting in 2020.  %that limits the achievable speedup of the Ethereum mainnet. 

This development presents a tremendous challenge in the quest to reach throughput levels of established payment systems. We, therefore, conclude by outlining three areas to tackle in order to increase the parallelizability of Ethereum's workload and allow concurrency mechanisms to reach their full potential. 
These areas should not only be targeted by Ethereum, the focus of our analysis, but also by blockchains with comparable smart contract designs and usage patterns, as we expect them to have similar bottlenecks.  %Further, the problem would only be amplified on UTxO-based blockchains due to the inherently sequential nature of UTxO consumption.

%We further point out that some of DeFi's most important smart contracts, ERC-20 tokens, and \emph{decentralized exchange (DEX)} routers introduce unnecessary dependencies in the transaction graph and, thereby, limit the parallelization potential. To simulate reality without these unnecessary dependencies, we propose and implement a disentanglement of the transaction graph. While we find that our suggested disentanglement increases the parallelizability of the Ethereum mainnet, the achievable speedup remains limited. Thus, our work shows that the parallelizability of Ethereum's workload is limited and, thereby, presents a tremendous challenge in the quest to reach throughput levels of established payment systems.

\section{Related Work}
Blockchain throughput has been one of the first topics of Bitcoin research, and many solutions have emerged to tackle the issue, e.g., layer 2, sharding. Here, we concentrate on those solutions that aim to parallelize the workload through concurrent execution. These works directly study the parallelization of the workload, whose bounds we quantify. 

Sergey and Hobor~\cite{sergey2017concurrent} are among the first to explore smart contract concurrency for parallel execution of blockchain transactions. They provide an analogy between smart contracts and concurrent programming. In the scheme introduced by Zhang and Zhang~\cite{zhang2018enabling}, miners use concurrency control techniques to pre-compute a serializable schedule that can be utilized by validators replaying the block. By employing a dependency graph based concurrency control technique, Amiri et al.~\cite{amiri2019parblockchain} find a valid schedule execution that allows for non-conflicting transactions to execute in parallel. Our work, on the other hand, explores the blockchain transaction dependency graph to quantify the existing real-world potential of parallelization.

Additionally, a recent line of work surrounding smart contracts' concurrency leverages speculative execution. Dickerson et al.~\cite{dickerson2017adding}, and Anjana et al.~\cite{anjana2022optsmart} pre-compute a serializable concurrent schedule for a block's transactions through speculative execution, while Gelashvili et al.~\cite{gelashvili2022block} propose Block-STM, a parallel execution engine that avoids pre-computation. An estimation of the potential concurrency of speculative execution by miners is offered by Saraph and Herlihy~\cite{saraph2019empirical}. Chen et al.~\cite{chen2021forerunner} take speculative execution to a new level by speculatively executing transactions that are waiting to be included in a block. In contrast, we explore the limits of concurrency given the nature of blockchain transactions in light of the recent rise of DeFi and NFT marketplaces.

Pîrlea et al.~\cite{pirlea2021practical} and Murgia et al.~\cite{murgia2021theory} utilize static analysis to parallelize execution. They statically determine which transactions can safely be executed in parallel and which contracts can be placed on different shards.  While static analysis is valuable to identifying dependencies ahead of time, the existing approaches do not remove inherent dependencies from the workload, which are at the center of our findings.
%Static analysis could be valuable for solving the problem of unnecessary dependencies, which we identify. On the other hand, the existing approaches are evaluated in practice and do not remove inherent dependencies from the workload, which are at the center of our findings.

A parallel line of work studies the transaction graphs of popular cryptocurrencies. Ron and Shamir~\cite{ron2013quantitative} first analyzed the Bitcoin transaction graph. While their work studies the full transaction graph statically, Kondor et al.~\cite{kondor2014rich} also examine changes in the Bitcoin transaction network over time. Several studies also explore the Ethereum transaction network~\cite{guo2019graph,motamed2019quantitative,motamed2019quantitative,chen2019dataether,bai2020evolution,ferretti2020ethereum,lin2020modeling,lin2021evolution,zhao2021temporal,he2021understanding,xie2021understanding,xie2021temporal,zhao2021temporal} through temporal graph analysis. Instead of solely studying the evolution of Ethereum's full transaction network, we focus on the impacts of the increased smart contract usage on the connectedness of Ethereum's block-wise transaction graphs and quantify the implications for parallelizability of the current and historical transaction workload.

In their study of Ethereum's transaction network, Zanelatto et al.~\cite{zanelatto2020transaction} focus on understanding the evolution of connected components in the network. Their work precedes the rise of DeFi and NFT marketplaces on the Ethereum and therefore does not capture the increased trend of more and more interplay between transactions and different smart contracts. Our work focuses on this increased connectivity and further discusses its implications on the blockchain.

\section{Background}

In the following, we introduce the essential preliminaries concerning transaction execution on the Ethereum blockchain and DeFi smart contracts. 

\subsection{Ethereum Transaction Execution}

Ethereum is a smart contract platform, i.e., it does not only support Ether transfers.
Instead, it runs a general purpose virtual machine, the \emph{Ethereum virtual machine (EVM)}, that executes a specific byte code instruction set. Thus, Ethereum allows functions defined in smart contracts to be called in transactions. Smart contract functions can call functions of other smart contracts, generating \emph{internal calls}. The EVM distinguishes between the following different types of function calls:~(1) \emph{call} performs operations scoped to the \emph{called} contract's storage,~(2) \emph{delegate call} performs operations scoped to the \emph{calling} contract's storage,~(3) \emph{static call} is like a regular call but with read-only access to the storage, and~(4) \emph{call code} is a deprecated version of delegate call.
Thus, all calls are scoped to a specific contract and have either read/write or read-only access to its storage.

%We call two transactions \emph{conflicting} or \emph{dependent} if they have calls with the same scope and one of them has write access on that scope.
%We will consider transactions conflicting if they interact with the same smart contract. Note that while two transactions that interact with the same smart contract might not necessarily touch the same storage cell, assuming them to be conflicting is in line the current information available for validators when building a block. While Ethereum introduced the \emph{access list}~\cite{2022eip2929,2022eip2930}, which specifies a list of addresses and storage keys that the transaction wants to access. The access list is optional and not widely adopted yet. Out of more than 600 million transaction only 2 million included an access list and their accuracy and completeness is unclear. Thus, validators building a parallelizable block have a higher chance of identifying the smart contracts a transaction interacts with than the specific storage keys, as these are even more dependent on the input parameters. Further, previous work could not gain a significant advantage at the memory level~\cite{saraph2019empirical} as opposed to the contract level.
We call two transactions \emph{conflicting} or \emph{dependent} if they have calls with the same scope and one of them has write access on that scope.
Note that two transactions that interact with the same smart contract might not necessarily touch the same storage cell. However, taking this coarser view on conflicts simplifies large scale analysis and is easier to reason about for smart contract developers.

There are also EVM instructions (BALANCE, EXTCODESIZE, EXTCODEHASH, EXTCODECOPY), which allow a contract to read global state (including the current Ether balance and smart contract code) of any address, regardless of the current scope.
In practice, out of these mostly EXTCODESIZE is used to detect whether an address is a smart contract. In our analysis we disregard potential conflicts, which could be caused by these instructions. 

Ethereum introduced the \emph{access list}~\cite{2022eip2929,2022eip2930}, which specifies a list of addresses and storage keys that the transaction wants to access, giving a gas discount on these accesses. However, it is optional and not yet widely adopted. We found that, out of more than 600 million transactions only 2 million included an access list and their accuracy and completeness is unclear. Building complete and accurate access lists of just addresses would also be easier for smart contract developers and end users. Further, previous work could not gain a significant advantage by parallelizing at the storage key level~\cite{saraph2019empirical} as opposed to the address level.

Code execution on the EVM is paid for in \emph{gas}, which is automatically converted from the sender's Ether balance. Gas is a measure for how expensive code execution is for validators on Ethereum~\cite{2022gas}.
Thus, we use gas as a proxy measure for real execution time (including computation and storage accesses).
%Note that previous work also uses gas to estimate execution time~\cite{saraph2019empirical}.
This way assumptions about the underlying implementation, runtime environment, and hardware are kept to a minimum.
Further, we define \emph{sequential gas} as the highest cumulative gas cost of any sequential execution in a parallel schedule.

\subsection{Decentralized Finance Smart Contracts}
DeFi offers many financial services from traditional finance. Instead of relying on intermediaries, DeFi utilizes smart contracts. We elaborate on the functionality of some of DeFi's most important smart contracts in the following. 

\textbf{ERC-20 Tokens} are smart contracts that implement fungible tokens, i.e., they can represent anything that can be owned and exchanged in integer quantities, and adhere to a specific interface standard, the ERC-20 (Ethereum Request for Comments 20)~\cite{2022erc20}. These implement at least a given set of nine functions, including \texttt{transfer} (transferring an amount of tokens from the callers address to a given address) and \texttt{balanceOf} (getting token balance of a given address).

\textbf{Routers} are implemented by DEXs as a central frontend to their trading interfaces. These routers are stateless, %, i.e., they do not hold any token balances. 
they simply specify via which \emph{liquidity pools}, trading venues for ERC-20 tokens, trades are routed. The router makes the calls to the liquidity pools according to a pre-defined route.

\section{Data Collection}

We run an Erigon client~\cite{2022erigon} to collect Ethereum blockchain data. In particular, we query trace data for the whole blockchain history to better understand the parallelizability of the Ethereum mainnet workload, as well as trends over time. Trace data provides us the internal calls executed by each transaction. In the proceeding evaluation, we look at historical data by sampling 65 blocks per day at random over the whole history of Ethereum's mainnet blockchain. With the historical analysis we can observe long term trends in Ethereum's workload and parallelizability thereof. Additionally, we also look at recent data in more detail by considering every single block over the three-month period from 1 June 2022 through 31 August 2022. Through the recent data we hope to get an accurate view of the current state of parallelizability on the Ethereum blockchain. 

\section{Ethereum Mainnet Workload}

\begin{figure}[t]
\begin{subfigure}[t]{0.525\linewidth}
  \centering    
  \includegraphics[scale = 0.45]{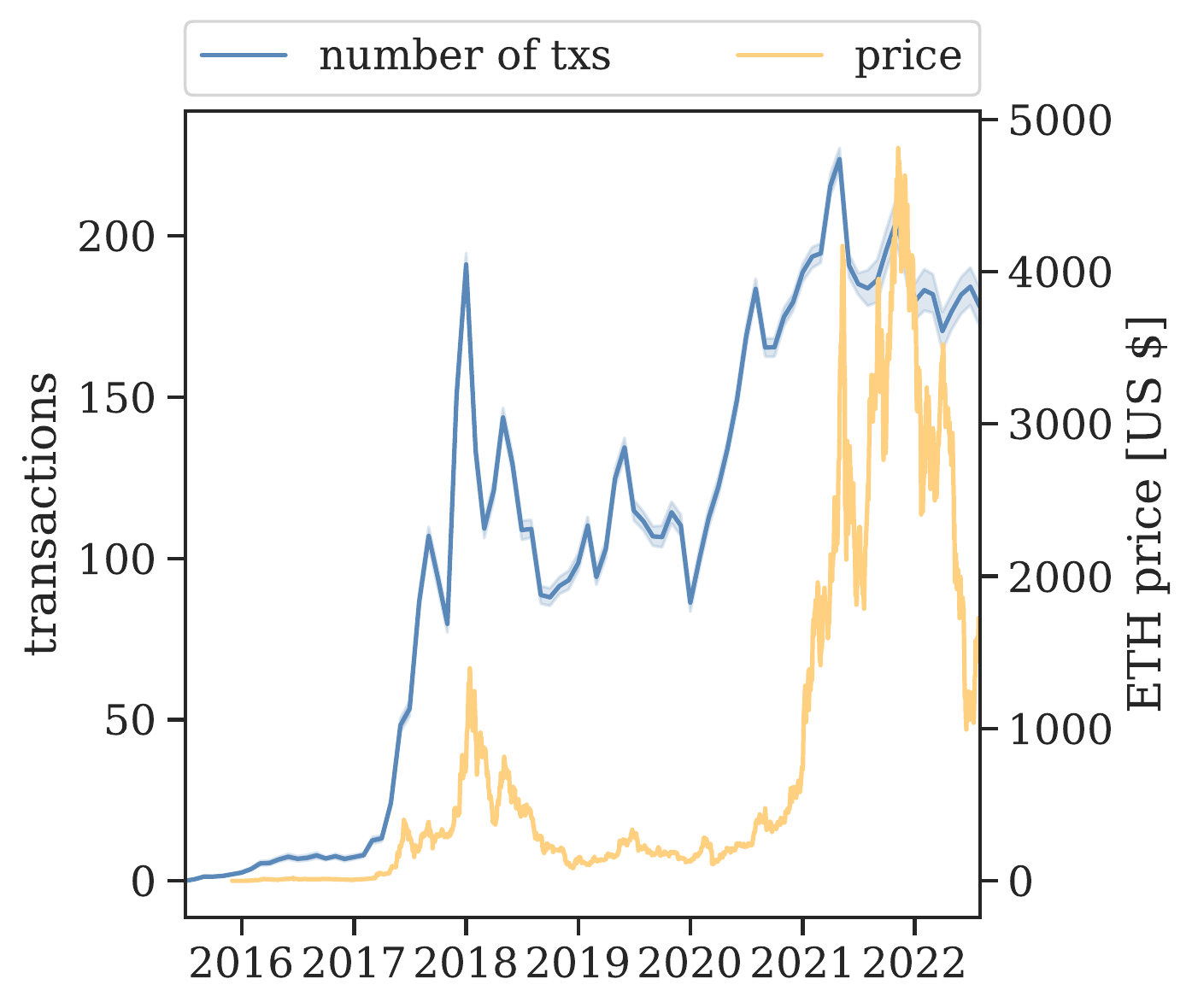}\vspace{-4pt}
  \caption{Historical development of the number of transactions per block on Ethereum mainnet and the Ether price.}
  %0.6851130941337945 
  %0.5983650694649768 after DeFi 
    %0.766017246459224 before DeFi
  %TODO data does not go until end of August
  \label{fig:num_txs}\vspace{-6pt}
  \end{subfigure}%
  \hfill
  \begin{subfigure}[t]{0.45\linewidth}
  \centering    
  \includegraphics[scale = 0.45]{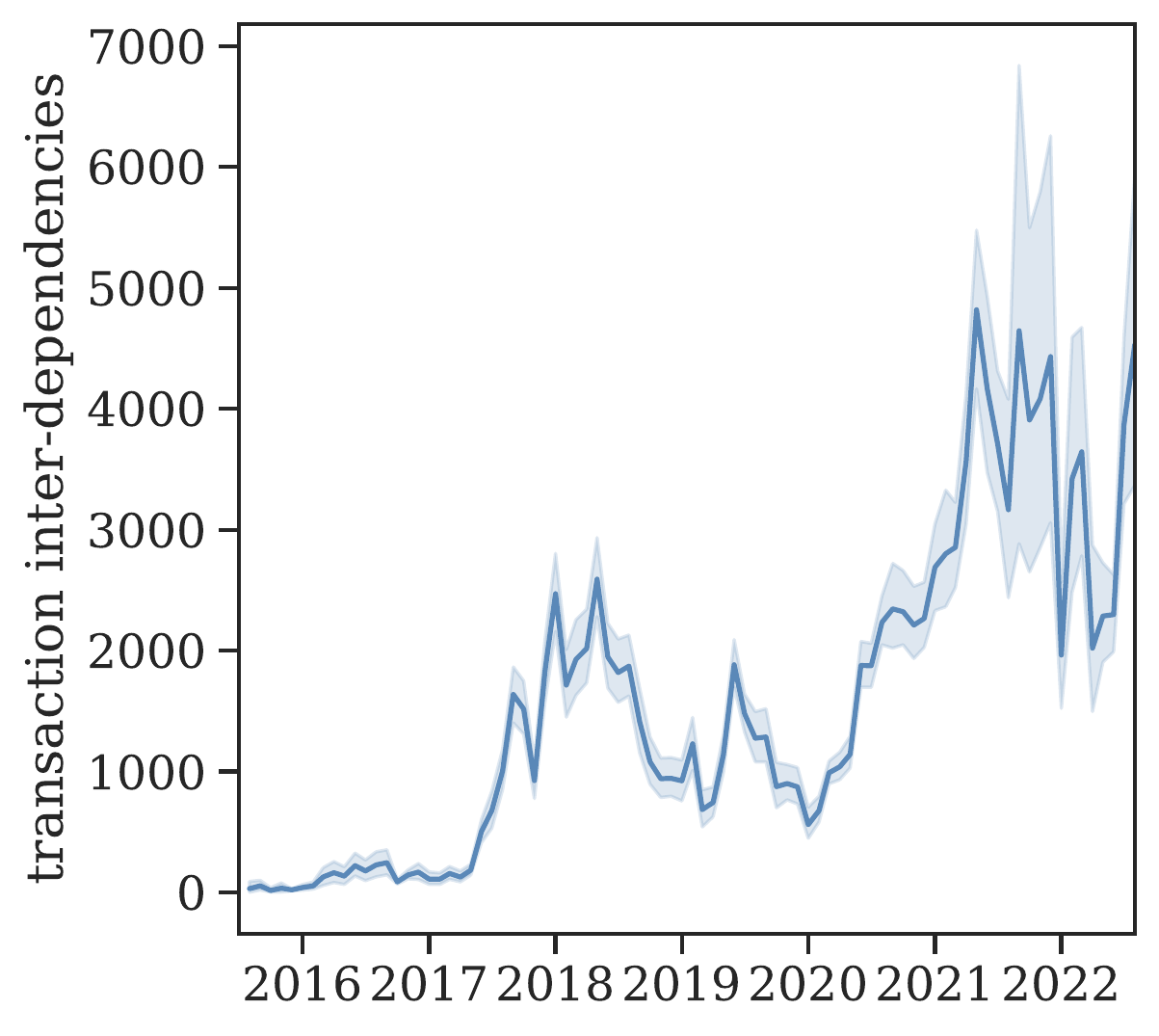}\vspace{-4pt}
    \caption{Historical development of the number of transaction interdependencies per block.} \label{fig:dependencies}\vspace{-6pt}
\end{subfigure}
\caption{Visualization of the average number of transactions per block (cf. Fig.~\ref{fig:num_txs}) and the average number of transaction interdependencies per block (cf. Fig.~\ref{fig:dependencies}). We randomly sample 65 blocks per day and plot the daily average along with the 95\% confidence interval. 
%Notice the sharp rise in the number of transactions in 2017, since then the number of transactions per block has not experienced a similar increase.
}
\end{figure}

In the following we consider changes in the Ethereum mainnet workload over time. In Fig.~\ref{fig:num_txs}, we compute and plot the average number of transactions of the blocks in our data set each month along with the 95\% confidence interval. Alongside this, we also show the Ether price.

While the number of transactions per block is initially only small, i.e., less than ten transactions, we notice a first significant rise in the number of transactions per block starting in early 2017 and peaking in late 2017 at around 200 transactions per block. During this time, there were unprecedented levels of speculation surrounding cryptocurrencies; this hype likely drove the rapid increase in the Ether price and the number of transactions on the Ethereum mainnet. The market's subsequent cool-down is reflected in a transaction number decrease to around 100 per block and a sharp Ether price decrease. In fact, until the end of 2019, the number of transactions per block and the Ether price are highly correlated, i.e., the Pearson correlation coefficient is 0.77. Later, another increase in the average transaction number occurred in 2020 during the DeFi and the subsequent NFT boom. Since then, the average number of transactions per block was stable at around 200, and the average number of transactions did not significantly surpass the previous peak. With the rise of DeFi and NFTs, the correlation between the number of transactions and the Ether price also decreased, i.e., the Pearson correlation coefficient drops to 0.60 from 2020 onward.

We further plot the number of transaction interdependencies in Fig.~\ref{fig:dependencies}, i.e., the number of transaction pairs that access the same smart contract. While the impact of DeFi and NFT marketplaces on the number of transactions does not significantly surpass previous levels, this does not hold for the number of transaction interdependencies. Notice that there were, on average, around 2000 transaction interdependencies per block during the initial peak. However, the increased usage of smart contracts starting from 2020 also increased the average number of transaction interdependencies per block to 5000 at its peak and never significantly dropped afterward. Note that these interdependencies are largely created by a few core DeFi smart contracts, i.e., popular ERC-20 tokens, DEX routers, and NFT marketplaces. Thus, the widespread adoption of these smart contracts presents a significant challenge to the parallelization of Ethereum transactions.

%\section{Methodology}
%In the following, we elaborate on the data collection and processing we perform in this analysis. In particular, we outline how we build Ethereum's network transaction graph and what simplification we perform to explore the limits of parallelizability.

\section{Transaction Graph Representation}%TODO go through adjust for NFTS
\begin{figure}[t]
\begin{subfigure}[t]{0.48\linewidth}
  
  \centering    
    \usetikzlibrary{arrows}
\thispagestyle{empty}
\makeatletter
\pgfkeys{/kiviat/label style/.style={align=left,anchor=180+360/\tkz@kiv@radial*\rang}}

\usetikzlibrary{decorations.pathreplacing, arrows, fit}
\definecolor{gray1}{rgb}{0.75, 0.75, 0.75}

\definecolor{color1}{HTML}{EA638C}
\definecolor{color2}{HTML}{5A88B8}
\definecolor{color3}{HTML}{573280}
\definecolor{color4}{HTML}{FFD080}
\definecolor{color5}{HTML}{417B5A}
\begin{tikzpicture}[scale = 1]

    \def \n {1.5}
	\def \radius {1.5cm}
	\def \margin {15} % margin in angles, depends on the radius

	\node[draw, thick,circle, minimum size = 0.6cm, fill=black!10,inner sep = 0] at (0,0) (A)  {\scriptsize A};
	\node[draw, thick,circle, minimum size = 0.6cm, fill=black!10,inner sep = 0] at (1,2) (B) {\scriptsize B};
	\node[draw, thick,circle, minimum size = 0.6cm, fill=black!10,inner sep = 0] at (-1,1) (C) {\scriptsize C};
	\node[draw, thick,circle, minimum size = 0.6cm, fill =black!10,inner sep = 0] at (-2,-0.5) (D) {\scriptsize D};
	\node[draw, thick,circle, minimum size = 0.6cm, fill =black!10,inner sep = 0] at (2.25,1) (E) {\scriptsize E};
	\node[draw, thick,circle, minimum size = 0.6cm, fill =black!10,inner sep = 0] at (1.5,-0.25) (F) {\scriptsize F};
	\node[draw, thick,circle, minimum size = 0.6cm, fill=black!10,inner sep = 0] at (-2,2) (G) {\scriptsize G};
	\node[draw, thick,circle, minimum size = 0.6cm, fill =black!10,inner sep = 0] at (-0.75,-1) (H) {\scriptsize H};
	
    %\node[draw, thick,circle, minimum size = 1.3cm, fill =color2!50] at (2,0) (B) {DAI};
    %\node[anchor=west](v6) at (-1.8,-3.5){Uniswap};
    %\node[anchor=west](v7) at (1.1,-3.5){Sushiswap};
    \path[-{Stealth[length=1.3mm, width=1mm]}, draw]
    
    (A) edge[ thick,color4, bend left = 0] node[sloped,anchor=south,auto=false, inner sep = 1pt] { \textcolor{black}{ \scriptsize $tx_3$}}(B)
    (D) edge[ thick,color4, bend left = 0] node[sloped,anchor=south,auto=false, inner sep = 1pt] { \textcolor{black}{ \scriptsize $tx_3$}}(C)
    (B) edge[ thick,color3,dashdotted, bend left = 0] node[sloped,anchor=south,auto=false, inner sep = 1pt] {\textcolor{black}{ \scriptsize $tx_2$}}(C)
    (C) edge[ thick, color1, densely dashed,bend left ] node[sloped,anchor=south,auto=false, inner sep = 1pt] {\textcolor{black}{ \scriptsize $tx_1$}}(A)
    (C) edge[ thick, color4,bend right ] node[sloped,anchor=south,auto=false, inner sep = 1pt] { \textcolor{black}{ \scriptsize $tx_3$}}(A)
    (G) edge[ thick, color1, densely dashed, bend left = 0] node[sloped,anchor=south,auto=false, inner sep = 1pt] {\textcolor{black}{ \scriptsize $tx_1$}}(C)
    (E) edge[ thick, color2,dotted, bend left = 0] node[sloped,anchor=south,auto=false, inner sep = 1pt] {\textcolor{black}{ \scriptsize $tx_5$}}(F)
    (H) edge[ thick, color5,loosely dashed, bend left = 0] node[sloped,anchor=south,auto=false, inner sep = 1pt] {\textcolor{black}{ \scriptsize $tx_4$}}(D);
    
    %\node[draw, thick,circle, minimum size = 1.5cm, fill =BTC] at ({360/\n *3 +90}:\radius) (v5)  {BTC};
    %\path[thick]    (-2.5,-3.5) edge	node {}	(-2,-3.5);
    %\path[thick,dashed]    (0.4,-3.5) edge	node {}	(0.9,-3.5);
    
    \end{tikzpicture}\vspace{-4pt}
    \caption{Address-based graph representation of a sample set of Ethereum transactions, where vertices are addresses (contracts or wallets) and edges are calls. The color and style of an edge indicates which transaction the call belongs to.} \label{fig:graph1}\vspace{-6pt}
  \end{subfigure}%
  \hfill
  \begin{subfigure}[t]{0.48\linewidth}
  \centering    
    \usetikzlibrary{arrows}
\thispagestyle{empty}
\makeatletter
\pgfkeys{/kiviat/label style/.style={align=left,anchor=180+360/\tkz@kiv@radial*\rang}}

\usetikzlibrary{decorations.pathreplacing, arrows, fit}

\definecolor{color1}{HTML}{EA638C}
\definecolor{color2}{HTML}{5A88B8}
\definecolor{color3}{HTML}{573280}
\definecolor{color4}{HTML}{FFD080}
\definecolor{color5}{HTML}{417B5A}
\begin{tikzpicture}[scale = 1]

    \def \n {1.5}
	\def \radius {1.5cm}
	\def \margin {15} % margin in angles, depends on the radius

	\node[draw, thick,circle, minimum size = 0.6cm, fill =color4!50,inner sep = 0] at (0,0) (A)  {\scriptsize $tx_3$};
	\node[draw, thick,circle, minimum size = 0.6cm, fill =color1!50,inner sep = 0] at (-2,0) (B) {{ \scriptsize $tx_1$}};
	\node[draw, thick,circle, minimum size = 0.6cm, fill =color3!50,inner sep = 0] at (-1,1.5) (C) {\scriptsize $tx_2$};
	\node[draw, thick,circle, minimum size = 0.6cm, fill =color5!50,inner sep = 0] at (1.5,-0.5) (D) {\scriptsize $tx_4$};

	\node[draw, thick,circle, minimum size = 0.6cm, fill =color2!50,inner sep = 0] at (1.3,1.5) (G) {\scriptsize $tx_5$};
	
    %\node[draw, thick,circle, minimum size = 1.3cm, fill =color2!50] at (2,0) (B) {DAI};
    %\node[anchor=west](v6) at (-1.8,-3.5){Uniswap};
    %\node[anchor=west](v7) at (1.1,-3.5){Sushiswap};
    \path[ draw]
    
    (A) edge[ thick, bend left = 0] node[midway,below] {}(B)
    (B) edge[ thick, bend left = 0] node[midway,below] {}(C)
    (A) edge[ thick, bend left = 0] node[midway,below] {}(C)
    (A) edge[ thick, bend left = 0] node[midway,below] {}(D);
    
    %\node[draw, thick,circle, minimum size = 1.5cm, fill =BTC] at ({360/\n *3 +90}:\radius) (v5)  {BTC};
    %\path[thick]    (-2.5,-3.5) edge	node {}	(-2,-3.5);
    %\path[thick,dashed]    (0.4,-3.5) edge	node {}	(0.9,-3.5);
    
    \end{tikzpicture}\vspace{-4pt}
    \caption{Transaction-based graph representation of a sample set of Ethereum transactions, where vertices are transactions and the edges indicate dependency between two transactions. Transactions that interact with the same address are dependent.} \label{fig:graph2}\vspace{-6pt}
\end{subfigure}

\caption{Two types of graph representations of the same sample set of five Ethereum transactions. The edge colors indicate belonging to a transaction in the address-based graph representation (cf. Fig.~\ref{fig:graph1}), the transaction-based graph representation has the same transaction set as vertices.}
\end{figure}
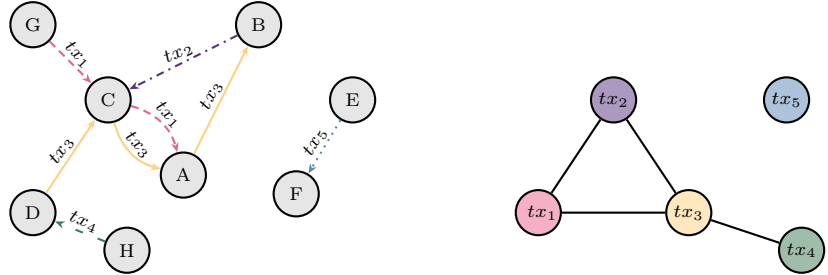

%TODO mention that not same storage used by all means
To explore trends in parallelization potential we consider a graph representation of the transaction data. There are two graph representations commonly utilized for Ethereum transaction data: \emph{address-based} and \emph{transaction-based}. We provide a definition for the address-based graph in Definition~\ref{def:ag} and visualize an example in Fig.~\ref{fig:graph1}. Observe that transaction $tx_3$ involves four address, $D\rightarrow C \rightarrow A \rightarrow B$, while $tx_2$ only involves two addresses, $B \rightarrow C$.

\begin{definition}[Address-based Graph (AG$_{n,m}$)] The address-based graph for blocks $n,\dots, m$, $n\leq m$, is represented as AG$_{n,m}(V,E,\{\omega_e\}_{e \in  E})$. Here, $V$ is the graph's set of vertices, each $v\in V$ is an address and $V$ is the set of all addresses that appeared in blocks $n,\dots, m$, i.e., were the sender or receiver of one of the (internal) calls of one of the block's transactions. $E$ is the graph's set of edges, each edge $e =(v,u)$ shows a call of contract $u$ by contract $v$ in blocks $n,\dots, m$. The weight of edge $e$, $\omega_e$, is given by the amount of gas utilized by the corresponding internal call.\label{def:ag}%TODO add footnote that there is a problem with weights not summing up
\end{definition}

In Definition~\ref{def:tg}, we define the transaction-based graph and draw the corresponding example in Fig.~\ref{fig:graph2}. Note that the address-based representation induces the transaction-based representation. In the transaction-based graph neighboring transactions cannot safely be executed in parallel. Therefore, in the example shown in Fig.~\ref{fig:graph2}, transaction $tx_4$ cannot be executed while $tx_3$ executes as they both interact with address $D$ (cf. Fig.~\ref{fig:graph1}). Cliques in the transaction-based representation indicate that all transactions in the clique have to be executed sequentially. Thus, the execution of transactions $tx_1$, $tx_2$ and $tx_3$ cannot be parallelized. In the following, we will largely consider the address-based representation, but will also draw unique insights from the transaction-based representation, i.e., calculate the graph's biggest clique to explore the limits of parallelization.

\begin{definition}[Transaction-based Graph (TG$_{n,m}$)] The transaction-based graph for blocks $n,\dots, m$, $n\leq m$, is represented as TG$_{n,m}(V,E,\{\omega_v\}_{v \in  V})$. Here, $V$ is the graph's set of vertices, each $v\in V$ is a transaction and $V$ is the set of all transactions that appeared in blocks $n,\dots, m$. The weight of a vertex $v$, $\omega_v$, is the amount of gas utilized by transaction $v$. $E$ is the graph's set of edges, each edge $e =\{v,u\}$ shows a dependency between transaction $v$ and $u$ in blocks $n,\dots, m$. A dependency is induced when the two transactions interact with the same address, i.e., the address appears as sender or receiver of an (internal) call for each of the two transactions.\label{def:tg}
\end{definition}

\begin{figure}[t]
\begin{subfigure}[t]{\linewidth}
  
  \centering    
    \includegraphics[width = 0.7\linewidth]{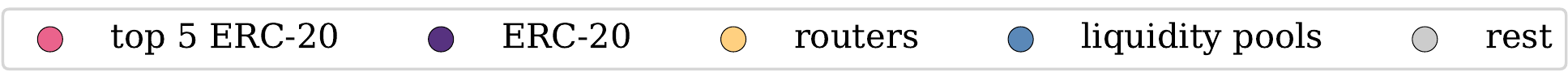}
  \end{subfigure}\vspace{2mm}

  \begin{subfigure}[t]{0.49\linewidth}
  
  \centering    
    \includegraphics[scale = 0.11]{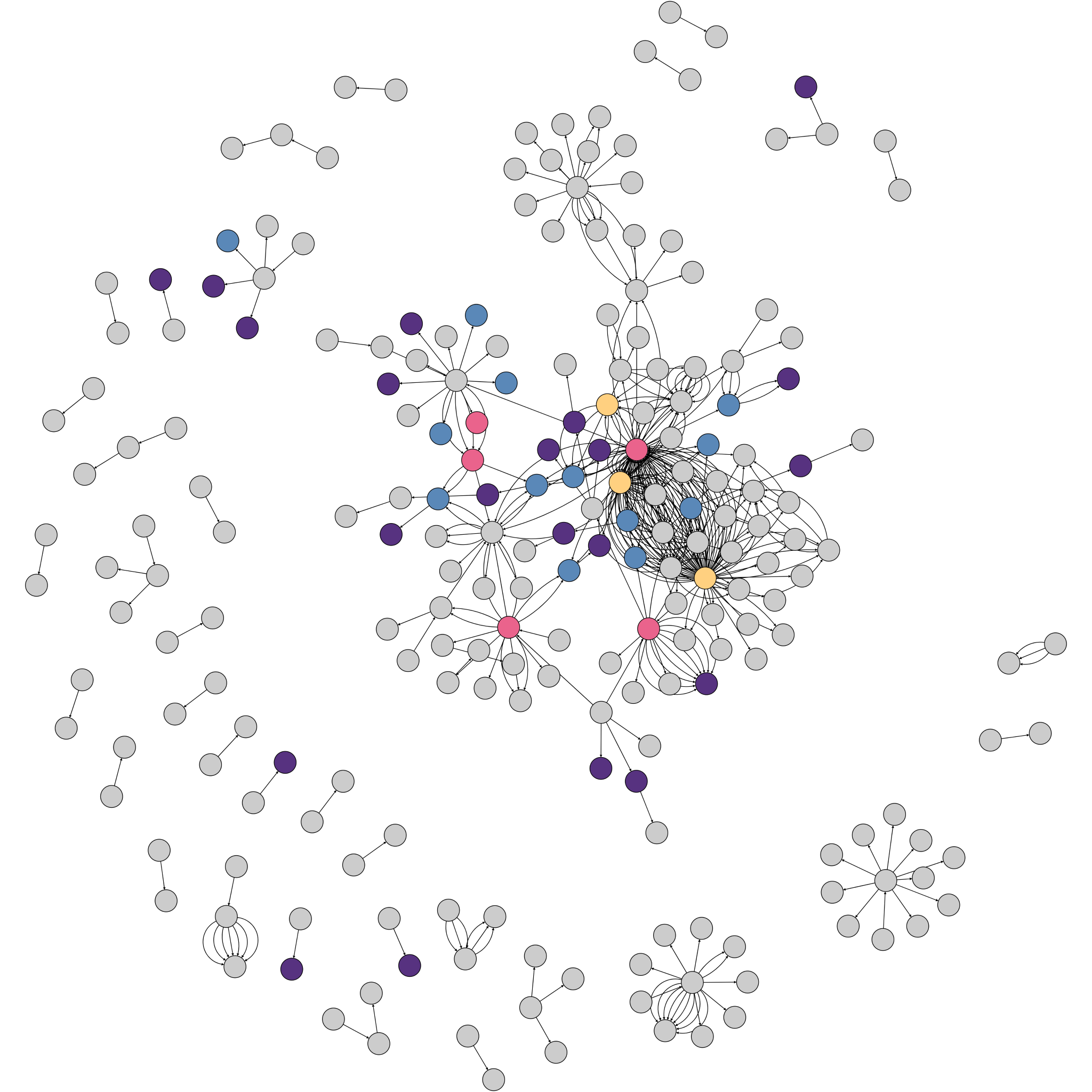}
    \caption{original address-based graph} \label{fig:blockorig}\vspace{-6pt}
  \end{subfigure}%34
  \hfill
  \begin{subfigure}[t]{0.49\linewidth}
  \centering    
    \includegraphics[scale = 0.11]{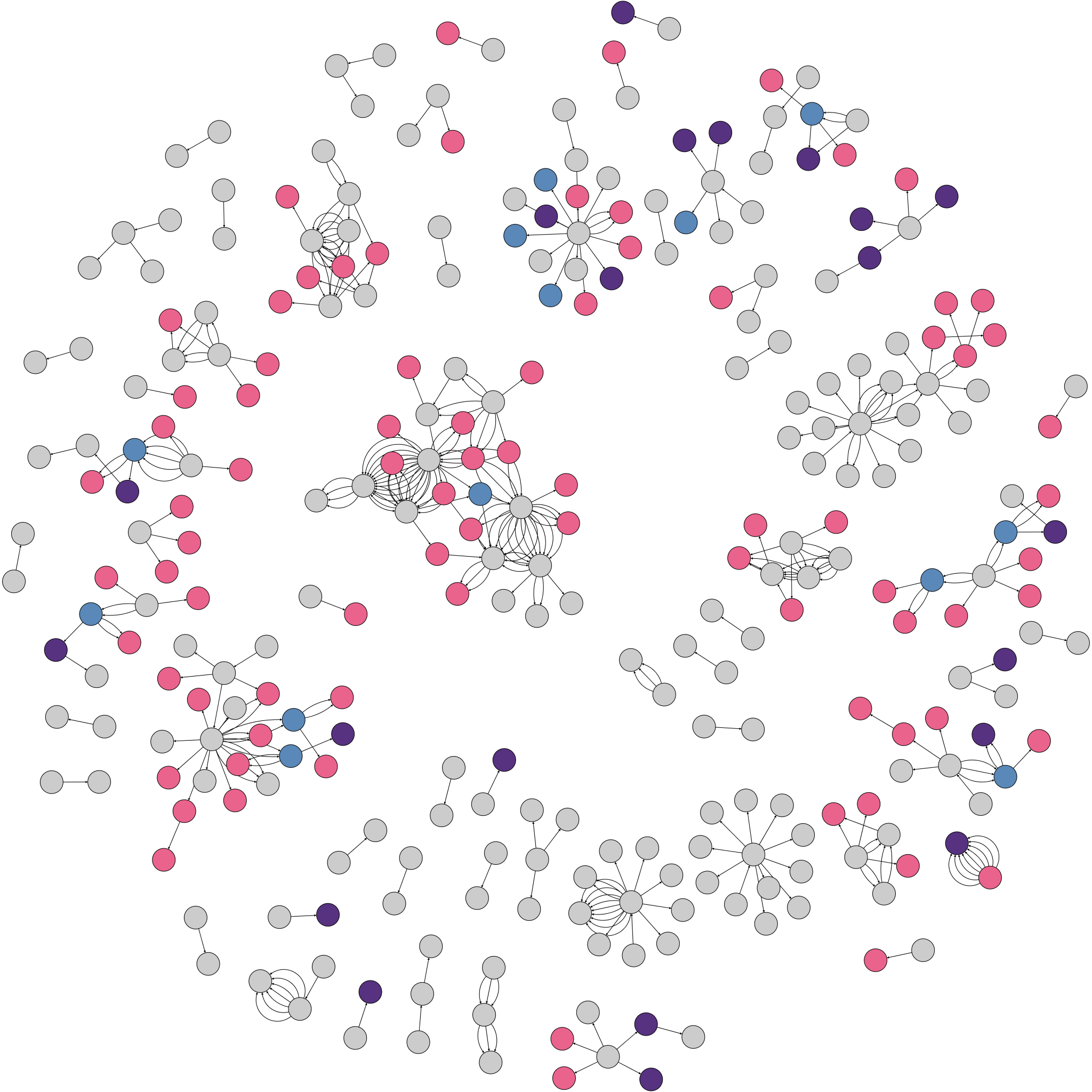}
    \caption{disentangled address-based graph}\label{fig:blockdisen}\vspace{-6pt}
\end{subfigure}

% \hfill
% \begin{subfigure}[t]{0.3\linewidth}
%   \centering    
%     \input{figures/transfers3.tikz}\vspace{0pt}
%     \caption{simplified address-based graph for ERC-20 transfer from} \label{fig:transfer3}\vspace{0pt}
% \end{subfigure}
\caption{Address-based visualisation of block 15,348,042 (mined 15 August 2022). Fig.~\ref{fig:blockorig} shows the original graph, while Fig.~\ref{fig:blockdisen} visualizes a disentangled version of the graph (cf. Section~\ref{sec:disen}). In both graphs we highlight some core DeFi smart contracts, namely, ERC-20 tokens, DEX routers, and DEX liquidity pools. Note that the biggest connected component of the disentangled graph is significantly smaller than that of the original graph.}
\end{figure}

In Fig.~\ref{fig:blockorig} we visualize the address interactions of block 15,348,042 (mined 15 August 2022) and highlight some core DeFi smart contracts. The pink vertices are the top five ERC-20 tokens (WETH, USDC, USDT, DAI, and LINK) in terms of the number of transfers, while the purple vertices are the remaining ERC-20 token addresses that appeared in block 15,348,042. We highlight the following DEX routers in yellow: Uniswap V2, Uniswap V3, SushiSwap, and 1inch, and utilize blue to flag the Uniswap V2,  Uniswap V3, SushiSwap, and Curve liquidity pools. Notice that the majority of the block builds a single connected component and that the vast majority of the labeled DeFi contracts are part of this connected component. With some, mainly the top 5 ERC-20 and the DEX routers, being central in this connected component and thereby contributing greatly to the dependencies between the different transactions in a block.

\subsection{Disentangled Transaction Graph Representation}\label{sec:disen}

Observing the persistently high degree (in the address-based transaction graph) of these DeFi contracts across the majority of the blocks since the rise of DeFi, we noticed that many of the dependencies introduced by these smart contracts, which are a central part of the DeFi ecosystem, are by no means essential. Especially the apparent dependencies introduced by ERC-20 token contracts and DEX routers in the smart contract level, would not manifest in the storage key level. These two examples of non-essential dependencies are relatively easy to spot by validators as we outline in the following. 
%These two examples of non-essential dependencies are relatively easy to spot by validators as we outline in the following.

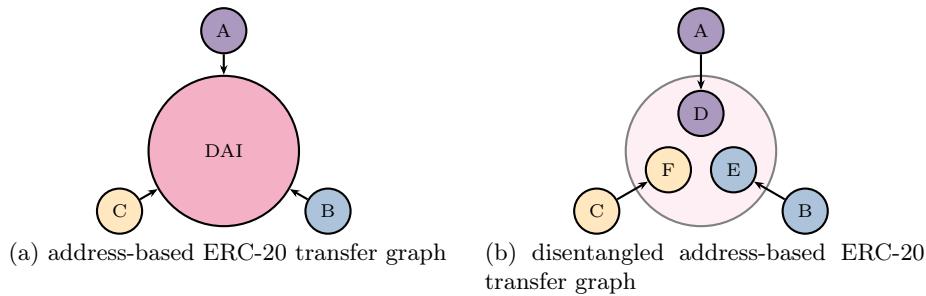
\begin{figure}[t]

\begin{subfigure}[t]{0.48\linewidth}
  
  \centering    
    \usetikzlibrary{arrows}
\thispagestyle{empty}
\makeatletter
\pgfkeys{/kiviat/label style/.style={align=left,anchor=180+360/\tkz@kiv@radial*\rang}}

\usetikzlibrary{decorations.pathreplacing, arrows, fit}

\definecolor{color1}{HTML}{EA638C}
\definecolor{color2}{HTML}{5A88B8}
\definecolor{color3}{HTML}{573280}
\definecolor{color4}{HTML}{FFD080}
\definecolor{color5}{HTML}{417B5A}
\begin{tikzpicture}[scale = 1]

    \def \n {1.5}
	\def \radius {1.6cm}
	\def \margin {15} % margin in angles, depends on the radius

	\node[draw, thick,circle, minimum size = 0.6cm, fill =color3!50,inner sep = 0] at ({360/\n *3 +90}:\radius) (A)  {\scriptsize A};
	\node[draw, thick,circle, minimum size = 0.6cm, fill =color2!50,inner sep = 0] at ({360/\n *1+90}:\radius) (B) {\scriptsize B};
	\node[draw, thick,circle, minimum size = 0.6cm, fill =color4!50,inner sep = 0] at ({360/\n *2+90}:\radius) (C) {\scriptsize C};
    \node[draw, thick,circle, minimum size = 2cm, fill =color1!50,inner sep = 0] at (0,0) (DAI) {\scriptsize DAI};
    %\node[draw, thick,circle, minimum size = 1.3cm, fill =color2!50] at (2,0) (B) {DAI};
    %\node[anchor=west](v6) at (-1.8,-3.5){Uniswap};
    %\node[anchor=west](v7) at (1.1,-3.5){Sushiswap};
    \path[-{Stealth[length=1.3mm, width=1mm]}, draw]
    
    (A) edge[ thick, bend left = 0] node[midway,below] {}(DAI)
    (B) edge[ thick, bend left = 0] node[midway,below] {}(DAI)
    (C) edge[ thick, bend left = 0] node[midway,below] {}(DAI);
    
    %\node[draw, thick,circle, minimum size = 1.5cm, fill =BTC] at ({360/\n *3 +90}:\radius) (v5)  {BTC};
    %\path[thick]    (-2.5,-3.5) edge	node {}	(-2,-3.5);
    %\path[thick,dashed]    (0.4,-3.5) edge	node {}	(0.9,-3.5);
    
    \end{tikzpicture}\vspace{-4pt}
    \caption{address-based ERC-20 transfer graph} \label{fig:transfer1}\vspace{-6pt}
  \end{subfigure}%
  \hfill
  \begin{subfigure}[t]{0.48\linewidth}
  \centering    
    \usetikzlibrary{arrows}
\thispagestyle{empty}
\makeatletter
\pgfkeys{/kiviat/label style/.style={align=left,anchor=180+360/\tkz@kiv@radial*\rang}}

\definecolor{color1}{HTML}{EA638C}
\definecolor{color2}{HTML}{5A88B8}
\definecolor{color3}{HTML}{573280}
\definecolor{color4}{HTML}{FFD080}
\definecolor{color5}{HTML}{417B5A}

\begin{tikzpicture}[scale = 1]

    \def \n {1.5}
	\def \radius {1.6cm}
	\def \margin {15} % margin in angles, depends on the radius

	\node[draw, thick,circle, minimum size = 0.6cm, fill =color3!50,inner sep = 0] at ({360/\n *3 +90}:\radius) (A)  {\scriptsize A};
	\node[draw, thick,circle, minimum size = 0.6cm, fill =color2!50,inner sep = 0] at ({360/\n *1+90}:\radius) (B) {\scriptsize B};
	\node[draw, thick,circle, minimum size = 0.6cm, fill =color4!50,inner sep = 0] at ({360/\n *2+90}:\radius) (C) {\scriptsize C};
	
    \node[draw,gray, thick,circle, minimum size = 2cm, fill =color1!10] at (0,0) (DAI) {};

    \def \radius {0.5cm}
	\node[draw, thick,circle, minimum size = 0.6cm, fill =color3!50,inner sep = 0] at ({360/\n *3 +90}:\radius) (DA)  {\scriptsize D};
	\node[draw, thick,circle, minimum size = 0.6cm, fill =color2!50,inner sep = 0] at ({360/\n *1+90}:\radius) (DB) {\scriptsize E};
	\node[draw, thick,circle, minimum size = 0.6cm, fill =color4!50,inner sep = 0] at ({360/\n *2+90}:\radius) (DC) {\scriptsize F};
    %\draw[ thick,fill=color3!50] (0,0) --  (150:0.6) arc(150:30:0.6) -- cycle;
    %\draw[ thick,fill=color4!50] (0,0) --  (225:0.6) arc(225:195:0.6) -- cycle;
    %\draw[ thick,fill=color2!50] (0,0) --  (345:0.6) arc(345:315:0.6) -- cycle;
    %\node[draw, thick,circle, minimum size = 1.3cm, fill =color2!50] at (2,0) (B) {DAI};
    %\node[anchor=west](v6) at (-1.8,-3.5){Uniswap};
    %\node[anchor=west](v7) at (1.1,-3.5){Sushiswap};
    \path[-{Stealth[length=1.3mm, width=1mm]}, draw]
    
    (A) edge[ thick, bend left = 0] node[midway,below] {}(DA)
    (B) edge[ thick, bend left = 0] node[midway,below] {}(DB)
    (C) edge[ thick, bend left = 0] node[midway,below] {}(DC);
    
    %\node[draw, thick,circle, minimum size = 1.5cm, fill =BTC] at ({360/\n *3 +90}:\radius) (v5)  {BTC};
    %\path[thick]    (-2.5,-3.5) edge	node {}	(-2,-3.5);
    %\path[thick,dashed]    (0.4,-3.5) edge	node {}	(0.9,-3.5);
    
    \end{tikzpicture}\vspace{-4pt}
    \caption{disentangled address-based ERC-20 transfer graph} \label{fig:transfer2}\vspace{-6pt}
\end{subfigure}
% \hfill
% \begin{subfigure}[t]{0.3\linewidth}
%   \centering    
%     \input{figures/transfers3.tikz}\vspace{0pt}
%     \caption{simplified address-based graph for ERC-20 transfer from} \label{fig:transfer3}\vspace{0pt}
% \end{subfigure}
\caption{Address-based graph representation of three transfers of DAI, an ERC-20 token, between the following wallets: address $A$ to address $D$, address $B$ to address $E$, and address $C$ to address $F$. In Fig.~\ref{fig:transfer1} we show the actual address-based representation. Observe that the three transfers appear dependent of each other. In Fig.~\ref{fig:transfer2} we show how we disentangle the graph to avoid this dependency.}
\end{figure}

For example, consider three transfers of DAI, an ERC-20 token, between the following addresses: address $A$ to address $D$, address $B$ to address $E$, and address $C$ to address $F$. As DAI is not the Ethereum blockchain's native currency, we only observe calls from the DAI senders to the DAI smart contract, which keeps track of fungible DAI tokens (cf. Fig.~\ref{fig:transfer1}). Thus, the three transfers appear dependent, which would not be the case for three equivalent ETH transfers. Therefore, we disentangled the transaction graph representation in Fig.~\ref{fig:transfer2}.
Instead of having the transaction's sender call the token contract, we pretend that they call the memory location of the receiver in the DAI smart contract. In addition to making these adjustments for the ERC-20 \texttt{transfer} function, we also make respective adjustments for the following ERC-20 contract functions: \texttt{balanceOf}, \texttt{transferFrom}, \texttt{approve}, and \texttt{allowance}.
Note that we only perform this disentanglement for the top five ERC-20 tokens: WETH, USDC, USDT, DAI, and LINK. These five ERC-20 tokens are the five largest in terms of the number of transfers and together account for 34\% of all ERC-20 transfers.
Further, we choose to restrict ourselves to this small number of ERC-20 tokens to show: (1) their impact on the connectedness of a block's transactions and (2) ensure that the tokens do not have any unexpected behavior, e.g., transferring a proportion to a third party~\cite{2022salmonella}.

DEX routers are also involved in many transactions and, thereby, lead to increased connectedness in the transaction graph. As the DEX routers themselves are stateless and only perform calls to the indicated liquidity pools on the user's behalf, the dependencies in the smart contract level are therefore not necessary. %In fact, the functionality of a router could be and often is implemented by the users themselves. 
Thus, we remove routers from the transaction graph. In particular, we re-route all the router edges to the sender of the respective transaction. In the later analysis, we perform this disentanglement for the routers of the following DEXs: Uniswap V2, Uniswap V3, SushiSwap, and 1inch.

%We want to note that the outlined disentanglement does not necessarily reflect the reality of the current Ethereum mainnet. Instead, it allows for an in-depth exploration of the potential for parallelization and paves the road for a possible overhaul of smart contract design to reduce dependencies. 

%Further, we propose a small number of simplifications to central parts of the DeFi ecosystem as well as a simple transaction scheduling algorithm and explore how much of the possible parallelizability can be achieved this way.

\section{Parallelizability}
In the following exploration of Ethereum's transaction graphs, we quantify the limited parallelization potential. In Section~\ref{sec:history}, we discuss the evolution over time -- cementing the impact of DeFi and NFTs -- and, in Section~\ref{sec:current}, the current state of parallelization potential on the Ethereum mainnet.
We define \emph{parallelizability} of a block as the highest speedup factor (total gas used by the block divided by sequential gas of a schedule) that can be achieved.
%under certain assumptions
For our analysis we look at specific schedules as well as graph metrics, which serve as upper and lower bounds on the parallelizability under our definition of conflicts.

\subsection{Parallelizability over Time}\label{sec:history}%TODO go through adjust for NFTs

In the following, we analyze the parallelization potential on Ethereum's mainnet by considering the connectedness of the transaction graphs. We randomly sample 65 blocks per day over the entire blockchain history up until the last block on 31 August 2022 -- allowing us to observe the trends over time.

\begin{figure}[t]
\begin{subfigure}[t]{0.48\linewidth}
  
  \centering    
    \includegraphics[scale = 0.45]{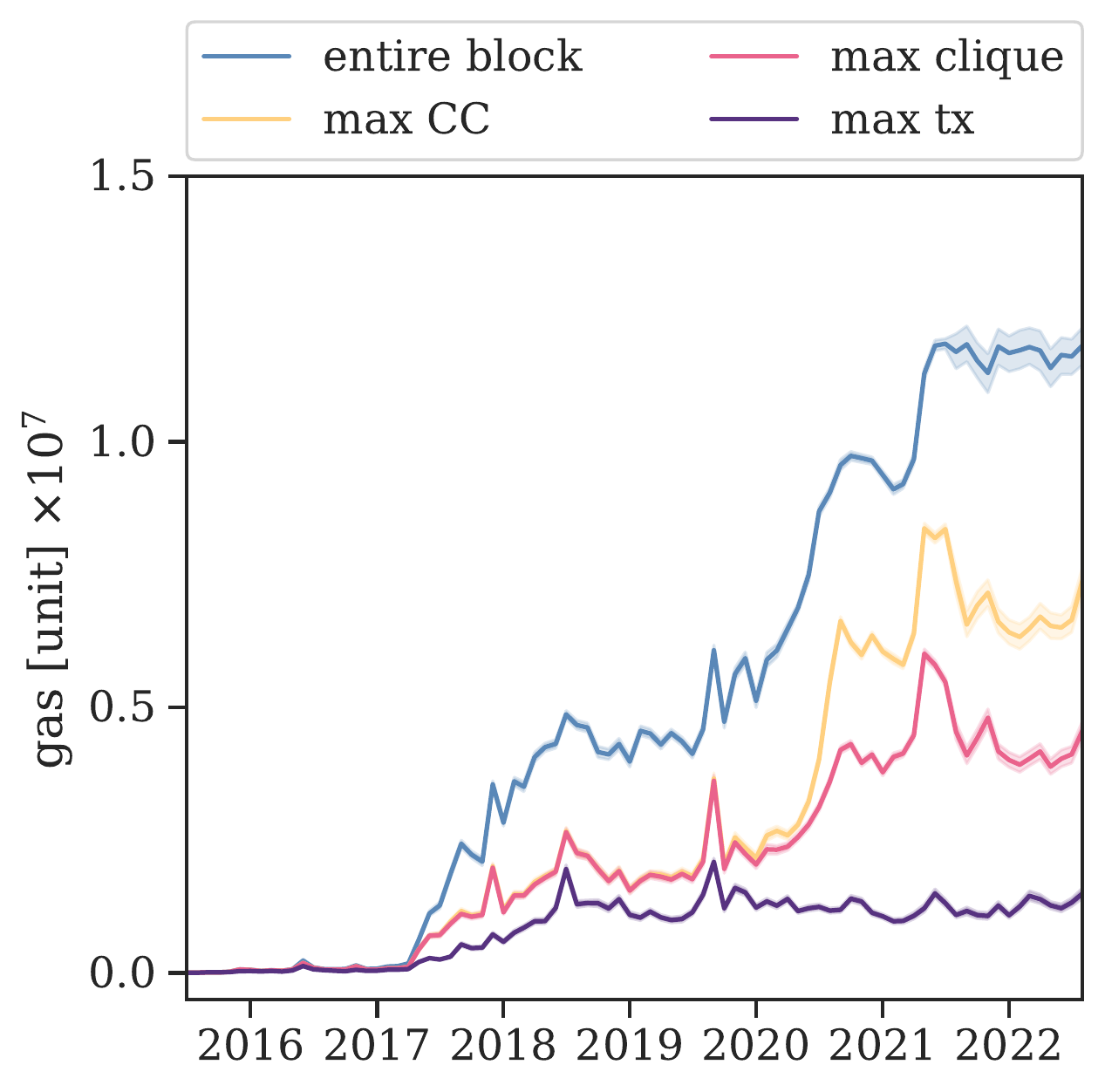}\vspace{-4pt}
    \caption{original transaction data} \label{fig:gashistory}\vspace{-6pt}
  \end{subfigure}%
  \hfill
  \begin{subfigure}[t]{0.48\linewidth}
  \centering    
    \includegraphics[scale = 0.45]{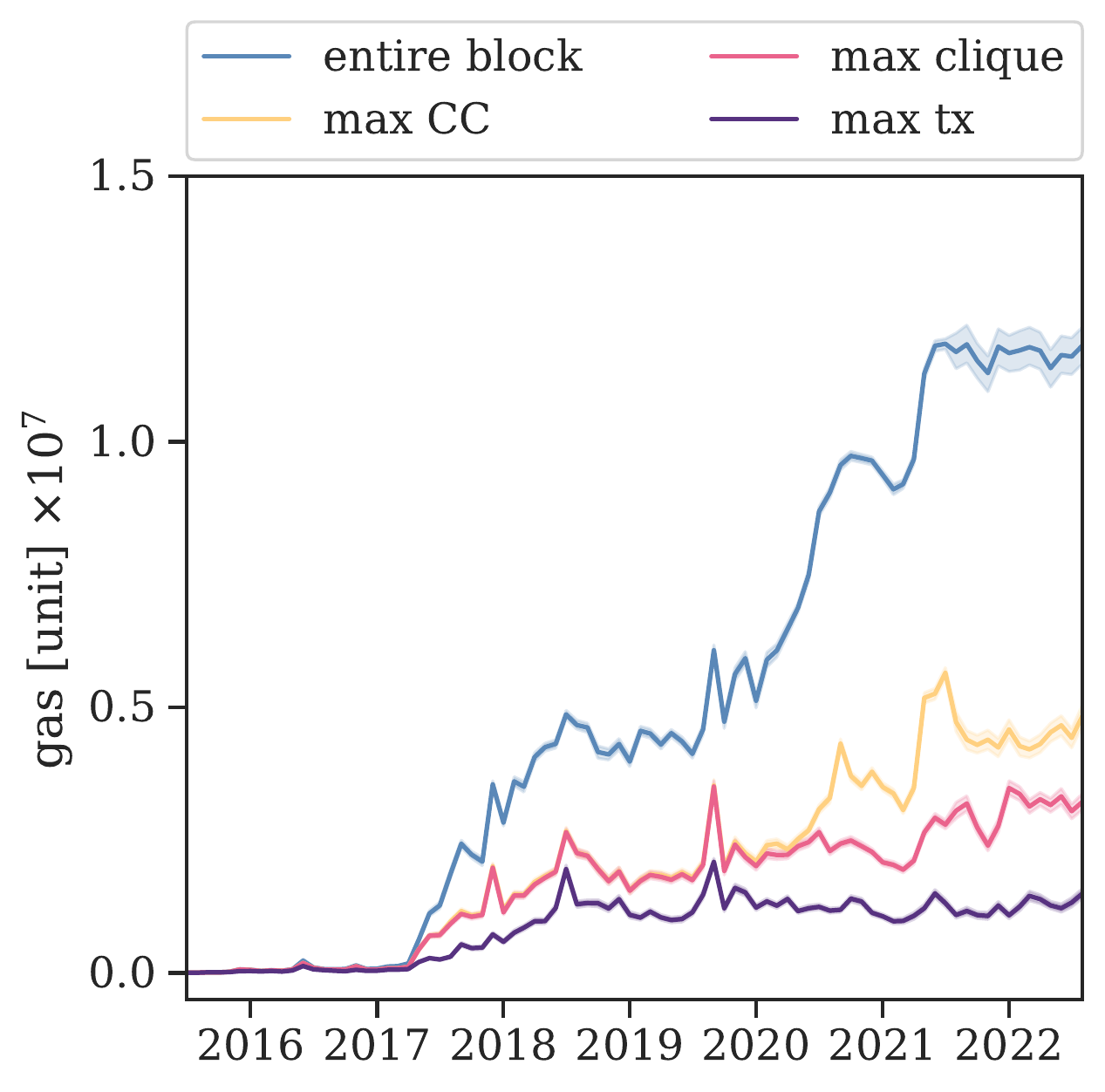}\vspace{-4pt}
    \caption{disentangled transaction data} \label{fig:gashistorysimplfied}\vspace{-6pt}
\end{subfigure}

\caption{We plot the gas used by: (1) the entire block, (2) the block's heaviest connected component (CC), (3) the block's heaviest clique, and (4) the block's heaviest transaction. Fig.~\ref{fig:gashistory} analyzes the original transaction data and Fig.~\ref{fig:gashistorysimplfied} the disentangled transaction data. Note that we plot the monthly average along with the 95\% confidence interval from randomly sampling 65 blocks per day.}
\end{figure}

The adoption of DeFi and NFT marketplaces is clearly visible when looking at trends over time in gas usage, a proxy for the execution time. In Fig.~\ref{fig:gashistory}, we plot the amount of gas per block in blue. Notice the sharp increase starting in 2020 with the rise of DeFi. Whereas these new applications did not increase the number of transactions to unprecedented levels, they caused the amount of gas per block to skyrocket due to the increasing complexity of transactions. 

To provide an enhanced understanding of the parallelizability of these increasingly heavy blocks, we also plot the size of the heaviest connected component, drawn in yellow, and the size of the heaviest clique, shown in pink, in Fig.~\ref{fig:gashistory}. Note that we measure the weight of a connected component or clique by the total amount of gas used by its transactions. Thereby, these weights indicate the time required to execute the contained transactions.

Connected components are equivalent in the address-based and transaction-based graph representations. Across both views, the weight of the heaviest connected component offers a lower bound for the parallelization potential of a block's execution. Any schedule that runs as many transactions as possible in parallel, i.e., in each time step executes a maximal independent set of transactions, will not exceed the time required to execute the heaviest connected component sequentially. We want to note that to obtain the dependencies between transactions, one has to have access to a statically provided access list or first execute all transactions. However, this is done only once by the validator. Once the block was executed, a parallel schedule could be made available to everyone else for validation.
Further, we utilize the transaction-based graph representation to find the heaviest clique. For this there is no direct analogue in the address-based graph. The heaviest clique specifies an upper bound for the parallelization potential of a block's execution. Any schedule must handle all transactions in a clique sequentially -- assuming atomic transaction execution. We, in fact, ran a simple list scheduling algorithm to find a schedule.
%Our algorithm finds a schedule that is sequentially consistent with the block's transaction ordering
It generates a partial ordering and always allows execution of a maximal independent set in parallel. We find that, while the schedule occasionally requires longer to execute than the heaviest clique would, the relative error is negligible (cf. Appendix~\ref{app:error}). Thus, the upper bound of the parallelization potential is almost achievable with a simple schedule. Note that our transaction graph might over estimate dependencies as we are coming from the smart contract level and not the storage key level. %Therefore, it is more of a realistic upper bound to parallelization given the information available to those building the block.

Looking at our data, we observe that the heaviest connected component currently makes up more than half of the block (cf. Fig.~\ref{fig:gashistory}). Further, the difference between the average heaviest connected component and the average heaviest clique has grown since the popularization of DeFi in 2020. This could be explained by interactions between the different protocols and smart contracts of the DeFi ecosystems. Since the rise of DeFi, the transactions in the heaviest connected component tend to interact with popular ERC-20 tokens, DEX liquidity pools, and lending protocols. However, mostly those that interact with the same smart contract are in a clique. We note that the largest clique typically consists of those transactions that interact with WETH, i.e., the most popular ERC-20 token (in terms of the number of transfers).
% TODO also mention NFTs here? (as a recent development)
% TODO what was the largest connected component before  ==> centralized exchanges, mining pools, early tokens & DEXs
%      over the whole history: WETH > USDT > USDC > Uniswap V2 Router > DAI (in CCs; cliques look similar)

As previously stated, transactions interacting with WETH generally form the heaviest clique in the original transaction data. However, as we show in Section~\ref{sec:disen}, the apparent dependencies 
%between these transactions are by no means necessary but rather 
in the smart contract level view are simply a consequence of implementing ERC-20 tokens as smart contracts as opposed to native tokens.  %TODO check whether we can say this
To obtain a more accurate picture, we perform the previously outlined disentanglement, we observe a significant reduction in both the size of the heaviest connected component and clique (cf. Fig.~\ref{fig:gashistorysimplfied}) since the adoption of DeFi. In fact, from 1 July 2020 to 31 August 2022, the disentanglement decreased the size of the heaviest connected component by a factor of 1.78 on average. Further, the size of the heaviest clique decreased by a factor of 1.88 on average.  

%Our suggested disentanglement demonstrates that not all apparent dependencies are necessary. Thus, 

We also plot the size, in terms of gas used, of the heaviest transaction per block in both Fig.~\ref{fig:gashistory} and Fig.~\ref{fig:gashistorysimplfied}.\footnote{Note that the disentanglement does not impact the size of the heaviest transaction, and neither the total gas of a block.} The size of the heaviest transaction in a block indicates a further, looser upper bound for the parallelization potential that disregards any dependencies between transactions. By neglecting all dependencies, we automatically omit any nonessential dependencies. This looser upper bound thus only assumes that transactions must execute atomically. However, we find that, on average, the heaviest transactions are a significant proportion of the entire block -- a ninth on average over the entire history. Thus, parallelization is not only limited by the ever-increasing size of the heaviest clique but is similarly bounded by individual large transactions. %TODO write why a transaction is a good unit

\begin{figure}[t]
\begin{subfigure}[t]{0.48\linewidth}
  
  \centering    
    \includegraphics[scale = 0.45]{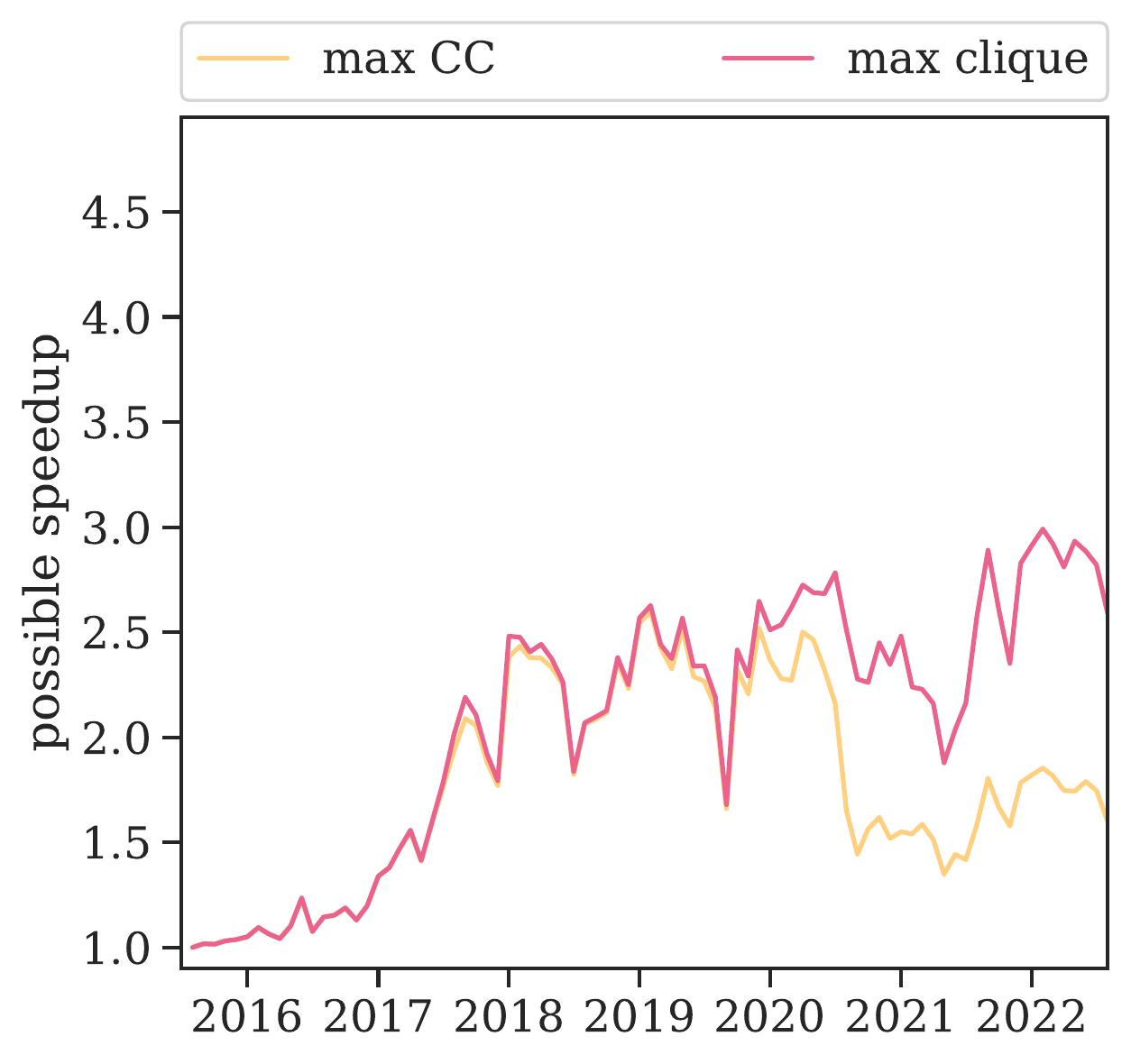}\vspace{-4pt}
    \caption{original transaction data} \label{fig:speeduphistory}\vspace{-6pt}
  \end{subfigure}%
  \hfill
  \begin{subfigure}[t]{0.48\linewidth}
  \centering    
    \includegraphics[scale = 0.45]{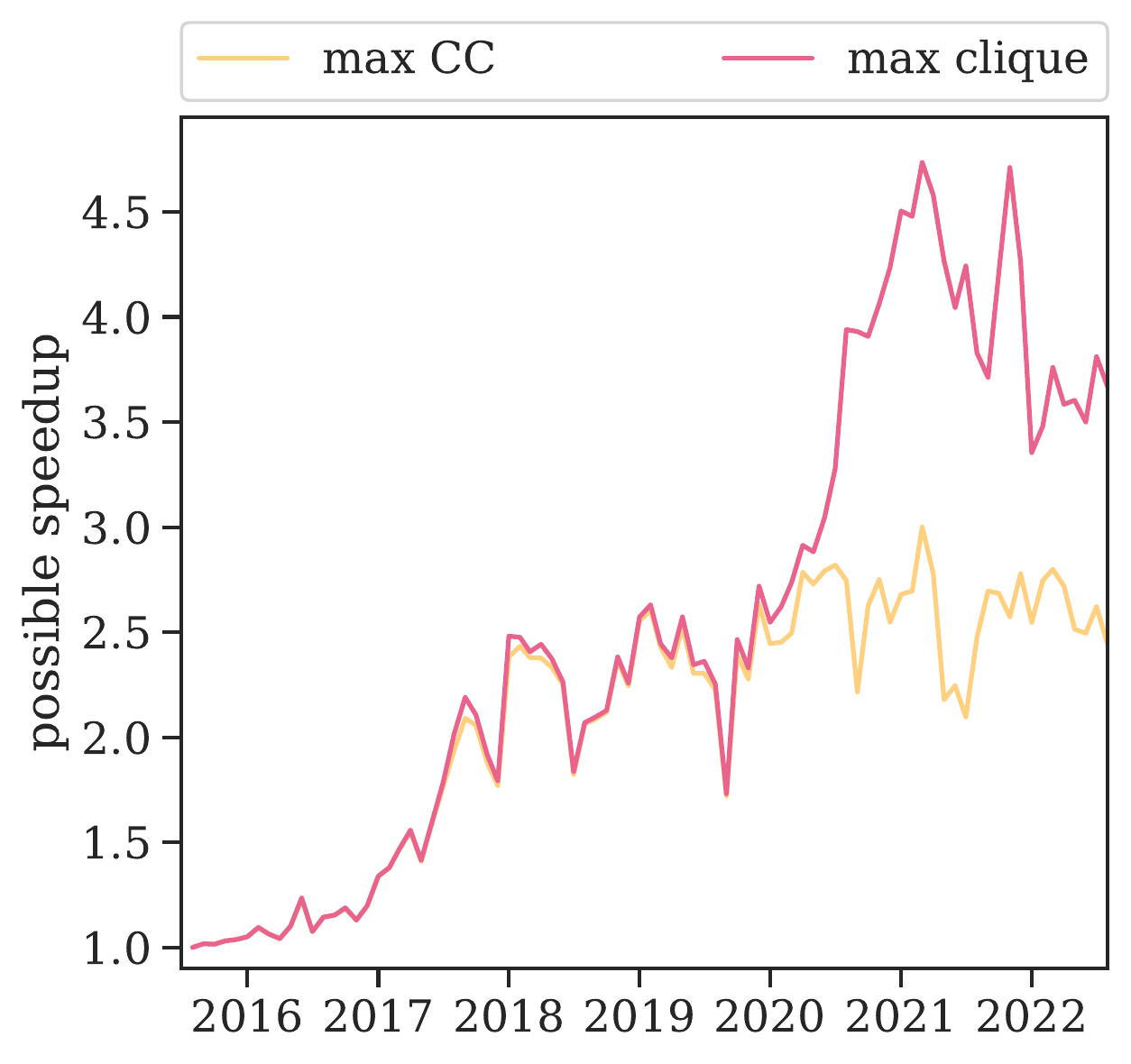}\vspace{-4pt}
    \caption{disentangled transaction data} \label{fig:speeduphistorysimplfied}\vspace{-6pt}
\end{subfigure}
\caption{We visualize the achievable execution speedup (aggregated monthly) through parallelization for the original transaction data (cf. Fig.~\ref{fig:speeduphistory}) and the disentangled transaction data (cf. Fig.~\ref{fig:speeduphistorysimplfied}). We obtain the lower bound for the parallelization potential through the identification of the heaviest connected component and the upper bound from the heaviest clique. Note both the heaviest connected component and clique are weighted by gas.} \label{fig:speeduphistoryfig}
\end{figure}

We plot the lower and (realistic) upper bound for the achievable speedup in Fig.~\ref{fig:speeduphistoryfig}.
In Fig.~\ref{fig:speeduphistory}, we show these bounds for the original transaction data, and in Fig.~\ref{fig:speeduphistorysimplfied}, we show the same bounds for the disentangled transaction data.
Similar to our previous observations, the lower (given by the size of the heaviest connected component) and the upper bound (given by the size of the heaviest clique) of the realistically achievable speedup are close to each other up until the rise of DeFi in 2020.
Further, we observe the performed disentanglement also only shows its effects from 2020 onward, as it targets DeFi smart contracts.
From 2020, we notice an increase in the difference between the lower and upper bound of the achievable speedup.
Further, in the original transaction data, we observe that both the lower and the upper bound for the realistically achievable speedup decrease once DeFi becomes adopted (cf. Fig.~\ref{fig:speeduphistory}).
In the disentangled transaction data, on the other hand, we notice that the lower bound for the possible speedup does not decrease after the introduction of DeFi, but instead remains more or less constant (cf. Fig.~\ref{fig:speeduphistorysimplfied}).
It is even more remarkable that the upper bound for the realistically achievable speedup even increases after the introduction of DeFi for the disentangled transaction data.
We presume this stems from the increasing number of transactions in the same period (cf. Fig.~\ref{fig:num_txs}).
Further, it is likely impacted by most DeFi transactions being dependent on each other over a given number of hops in the graph representation but not necessarily being all in one clique. %-- especially with the performed disentanglement.
%Thus, our proposed disentanglement allows for the achievable speed up to increase rather than decrease given increased DeFi adoption. 

\subsection{Current Limits of Parallelizability}\label{sec:current}%TODO check whether schedule explanations make sense

\begin{figure}[t]
\begin{subfigure}[t]{0.48\linewidth}
  
  \centering    
    \includegraphics[scale = 0.45]{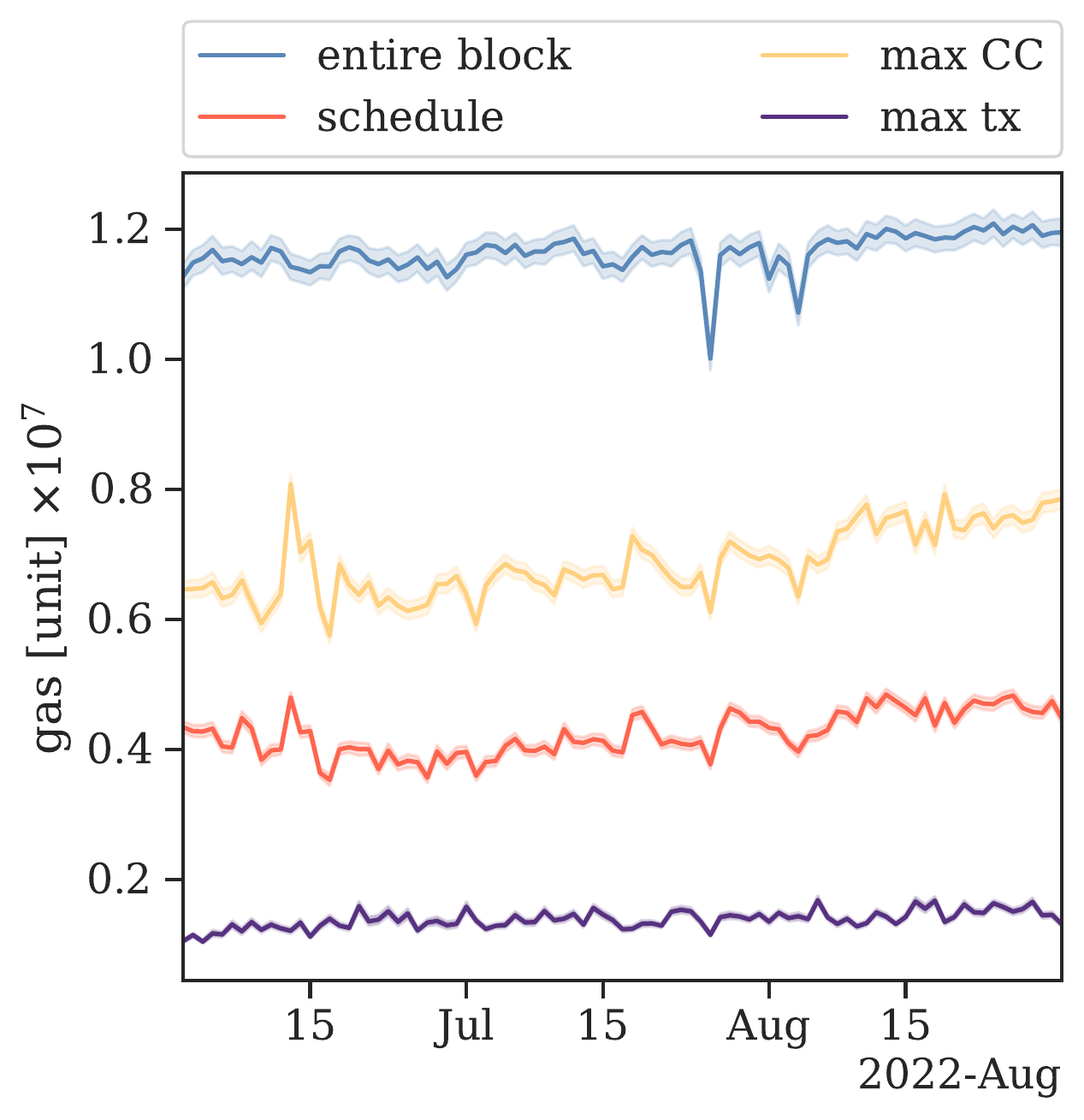}\vspace{-6pt}
    \caption{original transaction data} \label{fig:recentgas}\vspace{-6pt}
  \end{subfigure}%
  \hfill
  \begin{subfigure}[t]{0.48\linewidth}
  \centering    
    \includegraphics[scale = 0.45]{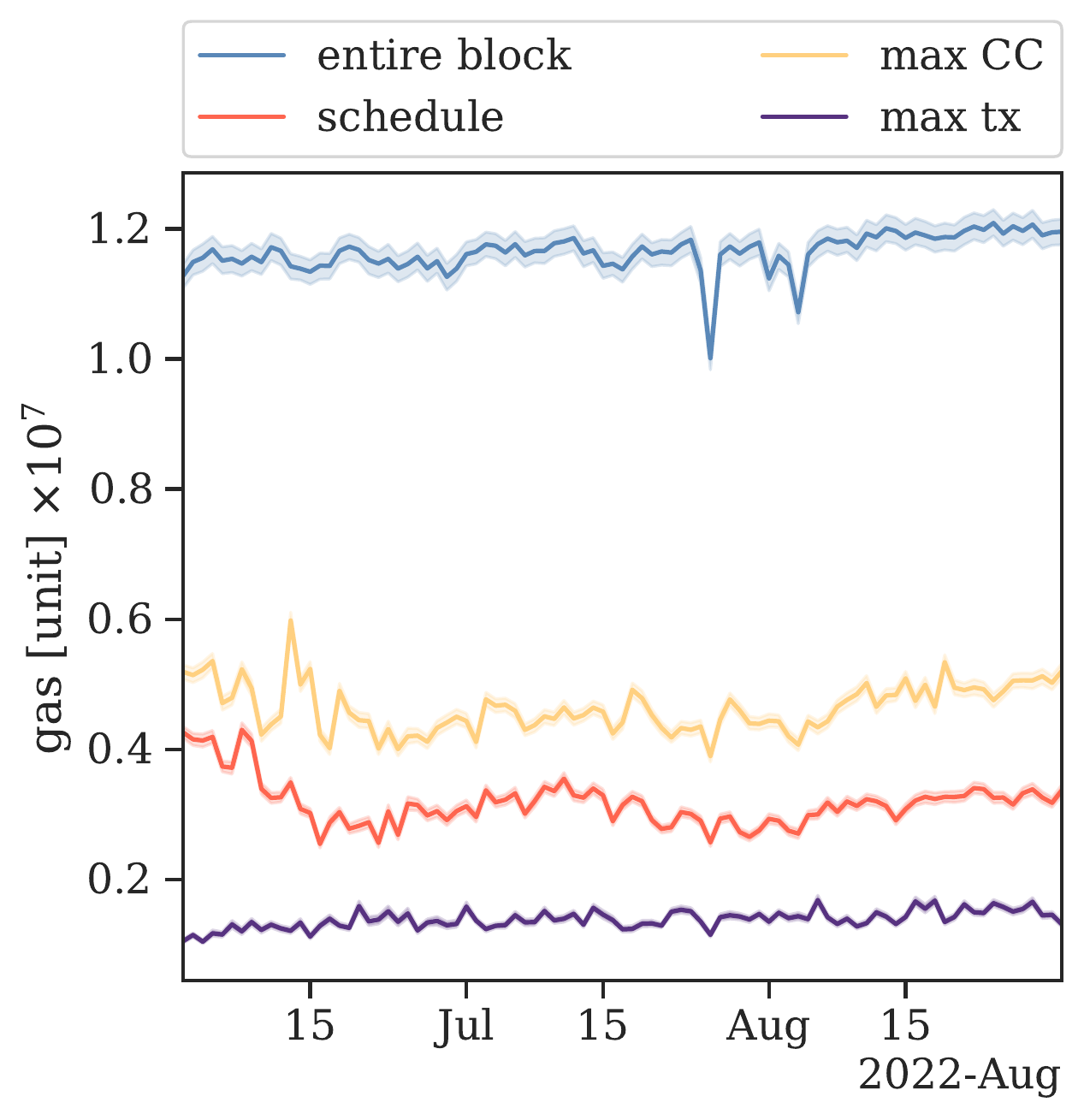}\vspace{-6pt}
    \caption{disentangled transaction data} \label{fig:recentgassimplfied}\vspace{-6pt}
\end{subfigure}
\caption{We plot the gas used by: (1) the entire block, (2) the block's heaviest connected component (CC), (3) our naive schedule (sequentially), and (4) the block's heaviest transaction. Fig.~\ref{fig:recentgas} analyzes the original transaction data and Fig.~\ref{fig:recentgassimplfied} the disentangled transaction data. Note that we plot the daily average along with the 95\% confidence interval. Further, we use the sequential gas of our schedule as a proxy for the size of the heaviest clique.} \label{fig:recentgasfig}%0.19036966248457962
\end{figure}

To better gauge the current limits of parallelizability, we expand on the previous analysis by analyzing all blocks from 1 June 2022 to 31 August 2022 -- allowing us to obtain a complete picture of the current state of the Ethereum mainnet. In Fig.~\ref{fig:recentgasfig}, we plot the amount of gas used by: (1) the entire block, (2) the block's heaviest connected component, (3) our schedule (sequentially), and (4) the block's heaviest transactions. Note that we only plot the sequential gas used by our schedule and not the heaviest clique, as finding the heaviest clique is time intensive. Our analysis in Appendix~\ref{app:error} shows that our schedule almost reaches the same parallelization potential. 

When examining Fig.~\ref{fig:recentgasfig}, we notice that there are few fluctuations in both the daily mean size of the blocks and the daily mean size of the heaviest transactions. We only observe two collapses, of around 10\%, in the mean size of the entire block at the end of July and the beginning of August. When looking at the daily average size of the heaviest connected component and the daily average amount of sequential gas used by our schedule in the original transaction data (cf Fig.~\ref{fig:recentgas}), a similar picture paints itself. In general, both averages make up approximately a half (connected component) and a third (schedule) of the block size on average. There is one peak in the average gas used by the heaviest connected component and the schedule around 15 June 2022 that we do not observe in the block size. The daily price movements of Ether were very high during this time (cf. Appendix~\ref{app:ethprice}, Fig.~\ref{fig:ethmove}) due to the anticipation of and the release of the CPI data~\cite{2022CPI}. As a consequence, the Ether trading volume on DEXs like Uniswap V3 experienced a rapid increase~\cite{2022UniEth}, which we presume lead to an increased size of both the heaviest connected component and clique in relation to the block size. We want to point out that, while the 95\% confidence interval is tight around the daily mean for all four graphs, the fluctuations of values for all four graph measures are substantial as shown in Appendix~\ref{app:recentgas} (cf. Fig.~\ref{fig:percentile}). For instance, shortly around the time at which we observe the peak in gas usage, the 99th percentile of the heaviest transaction reaches is almost the 99th percentile of the block size. Thus, there are some blocks in which a single transaction makes up almost the entire block -- allowing for little to no parallelization in those blocks. Still, we observe that the daily average of gas usage by the heaviest connected component and by the heaviest clique, for which we use our schedule as a proxy, make up a relatively stable proportion of the block in the recent (original) transaction data. 

Turning to the disentangled transaction data (cf. Fig.~\ref{fig:recentgassimplfied}), we notice a stable reduction in the daily average of the gas used by the heaviest connected components (by a factor of 1.6) and the sequential gas used by the schedule (by a factor of 1.5). It is most remarkable that the reduction is less significant in early June than in the remaining data set. We presume that this is a consequence of the significant price drop of Ether in the same period (cf. Fig.~\ref{fig:ethprice}), which likely led to exceptional DeFi usage patterns that further interconnected the workload. In the remaining data set, the reduction achieved by the disentanglement is very stable, but the achievable speedup still only reaches around a factor four (cf. Appendix~\ref{app:recentgas}, Fig.~\ref{fig:speeduprecentfig}).
\begin{figure}[t]
\begin{subfigure}[t]{0.48\linewidth}
  
  \centering    
    \includegraphics[scale = 0.45]{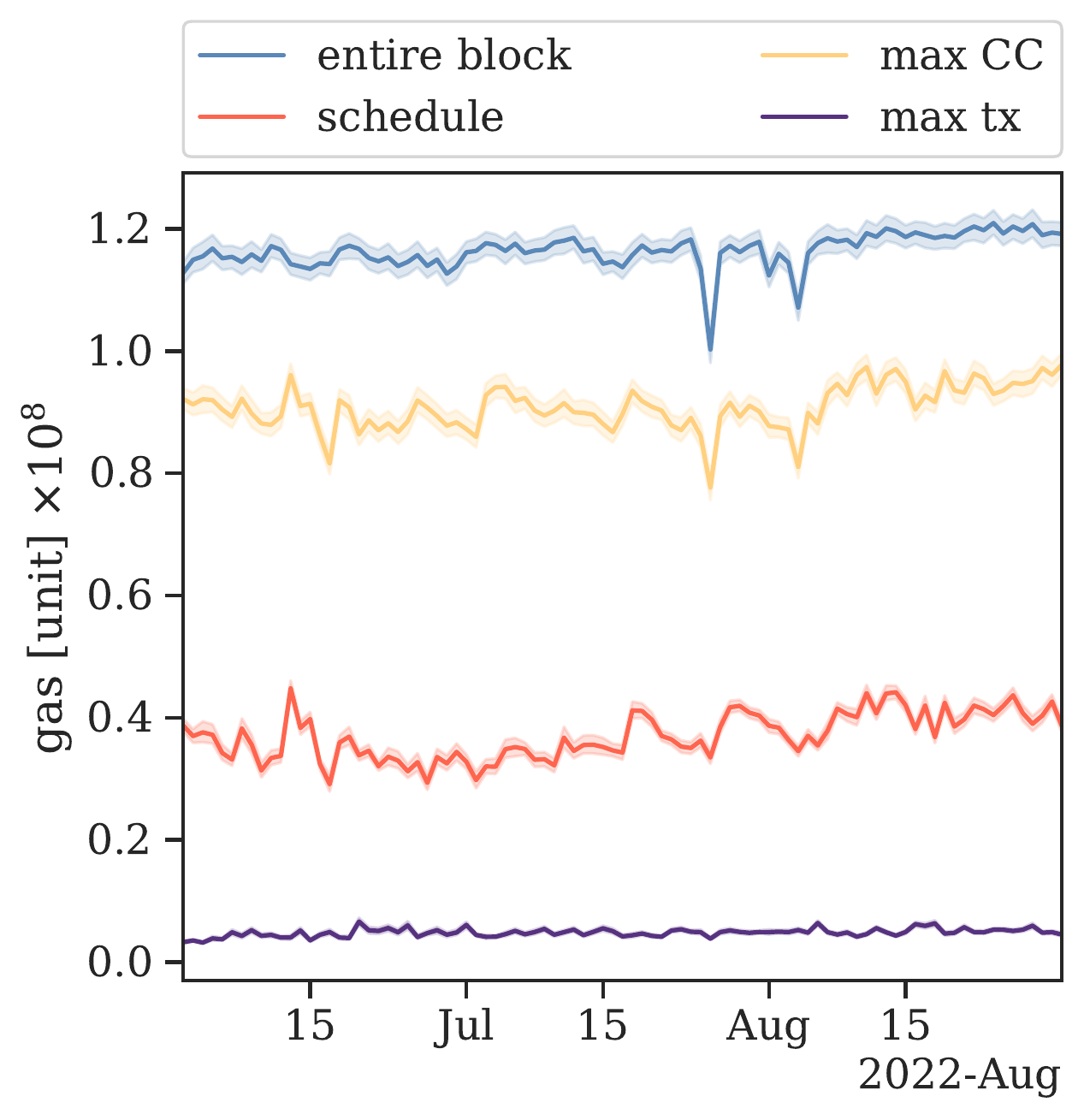}\vspace{-6pt}
    \caption{original transaction data} \label{fig:recentgas10}\vspace{-6pt}
  \end{subfigure}%
  \hfill
  \begin{subfigure}[t]{0.48\linewidth}
  \centering    
    \includegraphics[scale = 0.45]{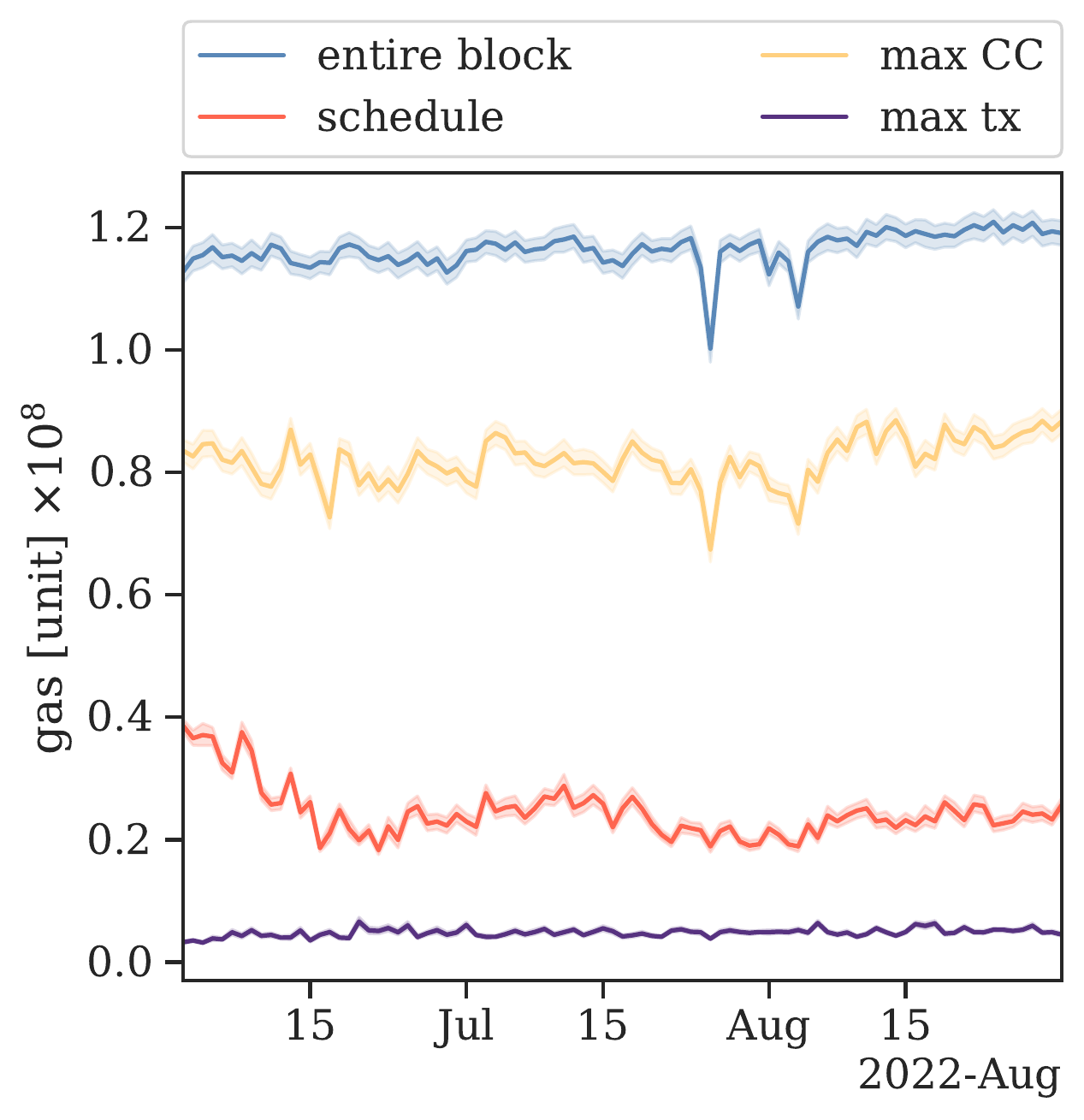}\vspace{-6pt}
    \caption{disentangled transaction data} \label{fig:recentgassimplfied10}\vspace{-6pt}
\end{subfigure}

\caption{We plot the gas used by: (1) 10 consecutive blocks, (2) their heaviest connected component (CC), (3) our naive schedule (sequentially), and (4) their  heaviest transaction. Fig.~\ref{fig:recentgas10} analyzes the original transaction data and Fig.~\ref{fig:recentgassimplfied10} the disentangled transaction data. Note that we plot the daily average along with the 95\% confidence interval.
%Further, we use the sequential gas utilized by our schedule as a proxy for the size of the heaviest clique.
} \label{fig:recentgasfig10}%0.19036966248457962
\end{figure}

Finally, in order to simulate higher transaction throughput, we consider batches of ten consecutive blocks and explore the connectedness of the corresponding transaction graphs (cf. Fig.~\ref{fig:recentgasfig10}). Even with this (rather exaggerated) simulated increase in block size, the sequential gas of our schedule increases proportionally, thus not changing this upper bound for the realistically achievable speedup. On the other hand, the lower bound (indicated by the heaviest connected component) even becomes much looser. This suggests that, when merging blocks, the largest cliques of all blocks merge into one, whereas connected components are even merged from within the same block. This is in line with our analysis that the heaviest cliques are always induced by the same few contracts -- indicating that increasing block size does not improve concurrency potential.

%Aiming to improve the understanding of the similarities and differences of the transaction network of various blockchains, Motamed and Bahrak~\cite{motamed2019quantitative} study the transaction graph of Bitcoin, Ethereum, Litecoin, Dash, and Z-Cash. They find that the growth rate of nodes and edges in the graphs is closely related to the price of the blockchain's native currency. In an Ethereum specific study, Bai et al.~\cite{bai2020evolution} study the evolutionary behavior of Ethereum transactions and similarly observe a correlation between the size of temporal user-to-user transaction graph and the Ether price.

\section{Discussion and Conclusion}

Our work quantifies the parallelizability of the Ethereum mainnet workload. We find that currently, the level of concurrency is very limited.
%It can by no means be parallelized efficiently on the 64 shards proposed in the Ethereum roadmap~\cite{2022shard}.
Thus, it does not suffice to only devote efforts to finding ways to best exploit the existing potential for concurrency. Instead, we believe that part of the focus must be shifted towards ensuring that the workload is parallelizable in the first place. Concretely, we believe that the following three areas must be targeted to enable the existing concurrency mechanisms to achieve much higher speedups.

%\textbf{Remove unnecessary dependencies.} As we outline, some of DeFi's core smart contracts introduce transaction dependencies that significantly hinder parallelization. Thus, we believe that there is potential for smart contracts to be redesigned to avoid any such unnecessary dependencies in the transaction graph. The incredibly low usage of the access list, indicates that it is currently not viable for transaction senders to provide the addresses and storage keys their transaction will touch. Predictability of dependencies would take care of this situation.  %TODO cite sui here??
%In case such dependencies no longer appear in the transaction graph, validators could more easily build a good execution schedule. 
%\textbf{Investigate dependencies.} As we outline, some of DeFi's core smart contracts introduce transaction dependencies that significantly hinder parallelization. Thus, we believe that the necessity of these dependencies must be investigated, and where possible, the smart contracts should be redesigned to avoid any such unnecessary dependencies in the transaction graph.
\textbf{Investigate dependencies.} As we outline, some of DeFi's core smart contracts appear in many transactions. Thus, we believe that transaction dependencies must be investigated on a more fine-grained basis, for example at the storage key level. Furthermore, the smart contracts could be redesigned to avoid unnecessary dependencies in the transaction graph.

\textbf{Incentivize ``simple'' transactions.} The heaviest transaction in a block currently makes up around one tenth of the average block size. Thus, these individual transactions present a limit on the parallelization potential. We therefore believe that the blockchain should discourage such frequent heavy transactions and instead encourage simple transactions. One possible approach would be charging for computation superlinearly.%\vspace{3pt}

\textbf{Increase predictability of dependencies.} The incredibly low usage of the access list, indicates that it is currently not viable for transaction senders to provide the addresses and storage keys their transaction will touch. Predictability of dependencies would take care of this situation and would allow for increased parallelization during execution.

Only once the workload on the Ethereum mainnet is truly parallelizable will the speedup suffice to make the 100,000 transactions per second~\cite{2022cryptonomist} stated by the Ethereum Foundation achievable.

%cannot be paralle. Thus, we believe that  and therefore calls for measures that change the transaction workload.

%Current efforts focus on how to best exploit the existing potential for concurrency. In the future, more work should be done investigating how to incentivize small and independent transactions. This might actually transform the transaction workload on the blockchain into a more parallelizable one and enable all of the existing concurrency mechanisms to achieve much higher speedups.

% smart contract design should be adapted, i.e.
% charge for 
% mention incentivizing "smaller" txs and (difficulty of) parallelizability of txs?

\bibliographystyle{splncs04}
\bibliography{references}
\newpage
% This creates an appendix chapter, comment if not needed.
\appendix
\section{Relative Error of Schedule}\label{app:error}
\begin{figure}[h]
\begin{subfigure}[t]{0.492\linewidth}
  \centering    
    \includegraphics[scale = 0.45]{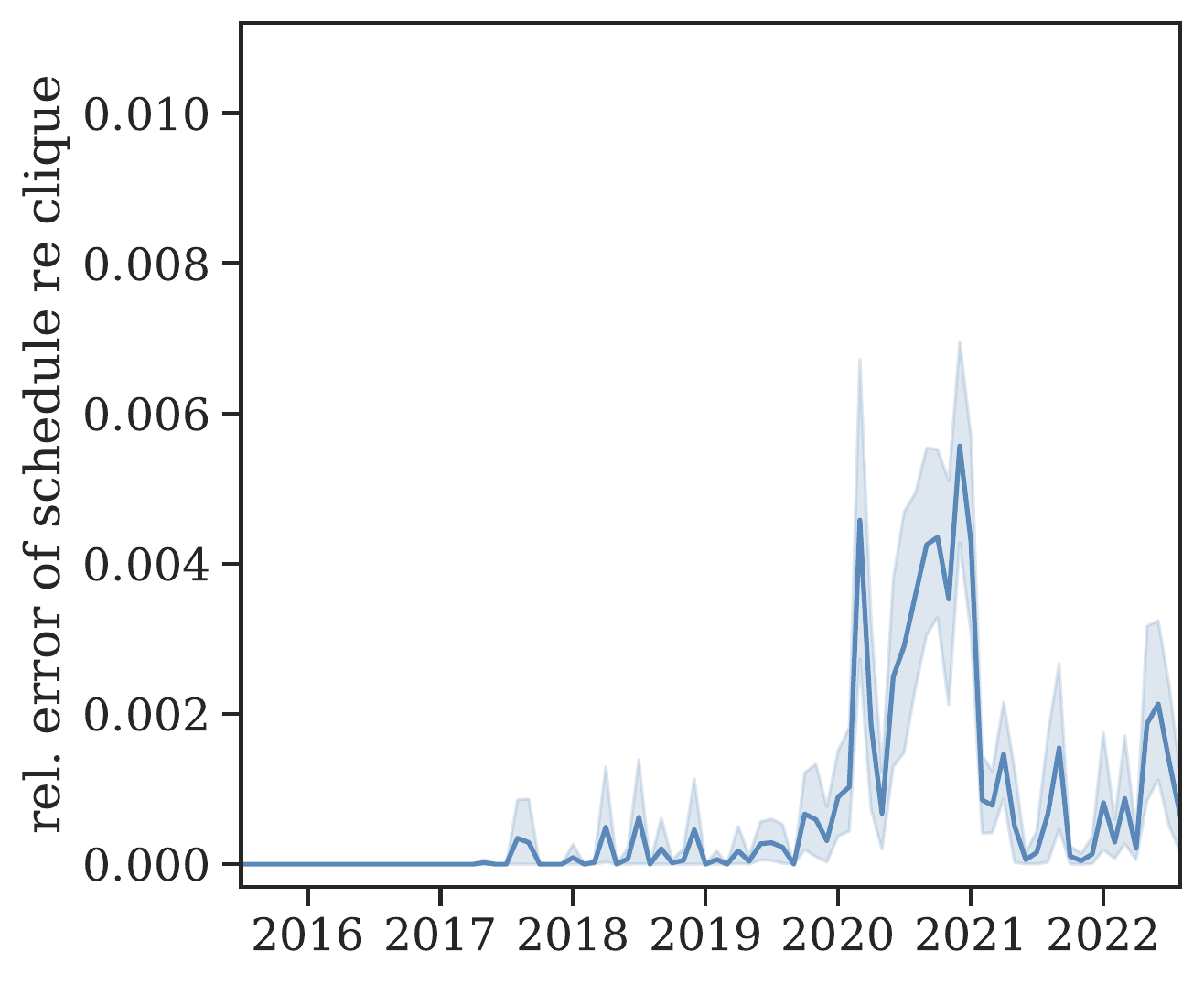}
    \caption{original transaction data} \label{fig:errororig}\vspace{0pt}
  \end{subfigure}%
  \hfill
  \begin{subfigure}[t]{0.492\linewidth}
  \centering    
    \includegraphics[scale = 0.45]{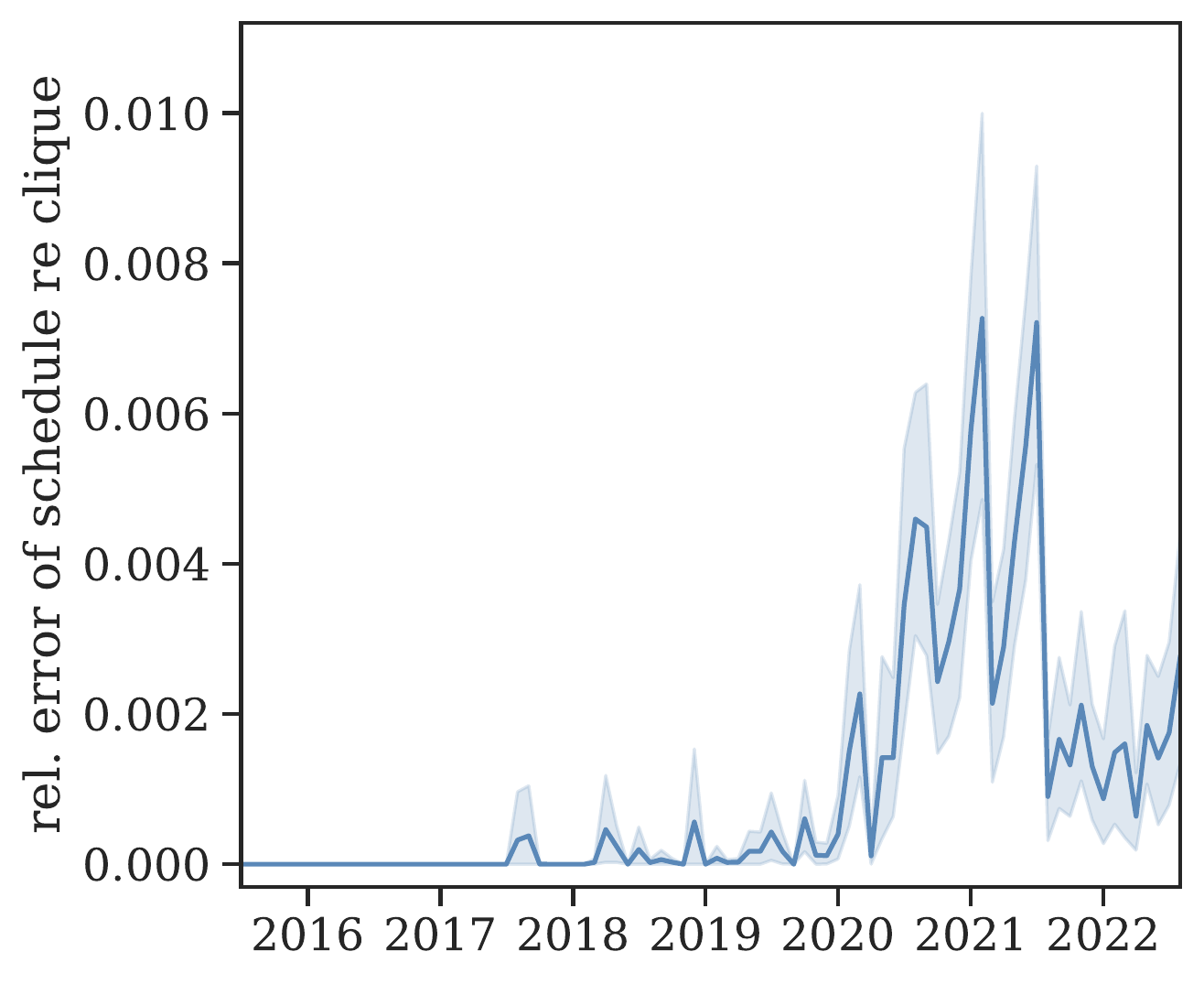}
    \caption{disentangled transaction data} \label{fig:errordis}\vspace{0pt}
\end{subfigure}
\caption{We visualize the relative error between the sequential gas of the schedule obtained through our naive scheduling algorithm and the heaviest clique, which is the best achievable for any parallel schedule working on the address level. We plot daily average relative error of 65 randomly sampled blocks per day and 95\% confidence interval for the original transaction data (cf. Fig.~\ref{fig:errororig}) and disentangled transaction data (cf. Fig.~\ref{fig:errordis}).}\label{fig:error}
\end{figure}

We compare the sequential gas used by our naive schedule to the amount of gas utilized for the execution of the heaviest clique in Fig.~\ref{fig:error}.
Notice that the mean daily relative error is very small, less than 1\%, throughout the entire blockchain history.
The relative error peaks for both the original transaction data and the disentangled transaction data, starting with the rise of DeFi up until mid-2021.
From mid-2021 onward, it becomes even smaller again.
We further point out that the relative error of our schedule is slightly larger for the disentangled transaction data than for the original data.
Regardless, the sequential gas utilized by our naive schedule accurately approximates the amount of gas utilized for the execution of the heaviest clique, which is optimal for any schedule working with conflicts on the granularity of addresses.

\section{Current Limits of Parallelizability}\label{app:recentgas}

\begin{figure}[t]
\begin{subfigure}[t]{0.48\linewidth}
  
  \centering    
    \includegraphics[scale = 0.45]{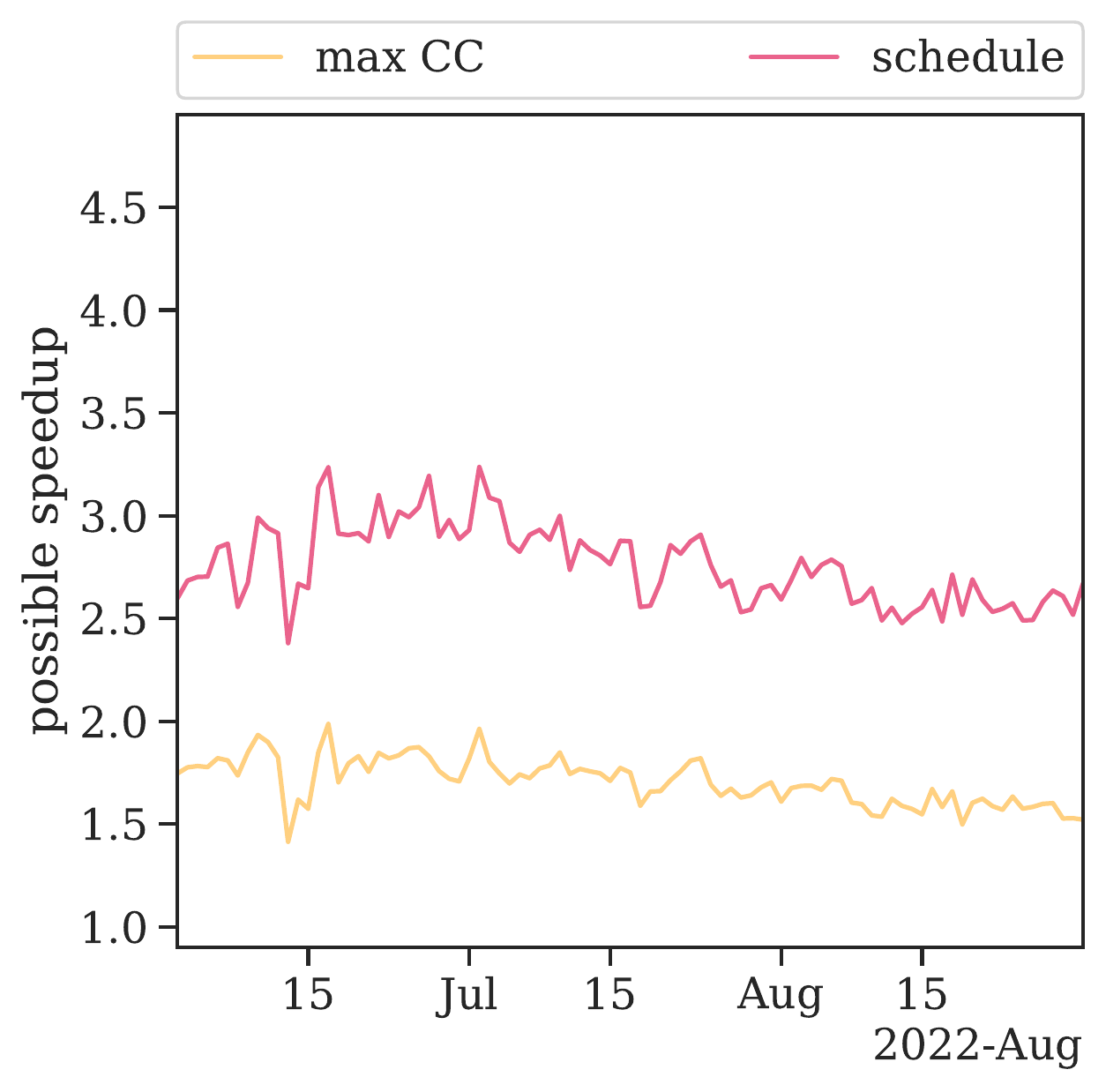}
    \caption{original transaction data} \label{fig:speeduprecent}\vspace{0pt}
  \end{subfigure}%
  \hfill
  \begin{subfigure}[t]{0.48\linewidth}
  \centering    
    \includegraphics[scale = 0.45]{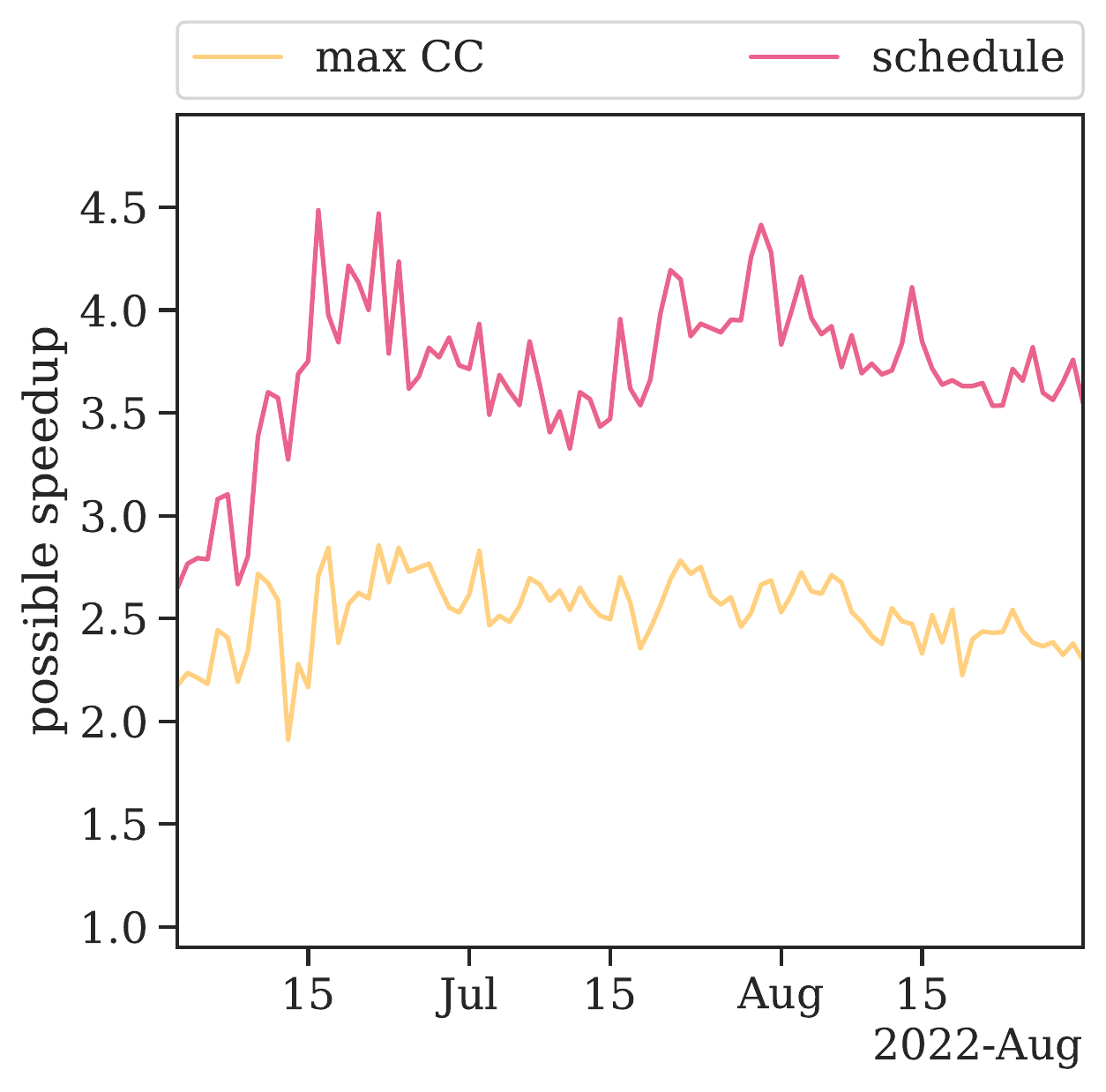}
    \caption{disentangled transaction data} \label{fig:speeduprecentsimplified}\vspace{0pt}
\end{subfigure}
\caption{We visualize the achievable execution speedup (aggregated daily) through parallelization for the original transaction data (cf. Fig.~\ref{fig:speeduprecent}) and the disentangled transaction data (cf. Fig.~\ref{fig:speeduprecentsimplified}). We obtain the lower bound for the parallelization potential through the identification of the heaviest connected component and the upper bound from the heaviest clique. Note that both the heaviest connected component and clique are weighted by gas.} \label{fig:speeduprecentfig}
\end{figure}

We plot bounds for the achievable execution speedup of the gas utilized for the execution of the entire block (averaged over each day), in Fig.~\ref{fig:speeduprecent} for the original transaction data and in Fig.~\ref{fig:speeduprecentsimplified} for the disentangled transaction data.
The heaviest connected component, in terms of gas used, serves as a lower bound, whereas the sequential gas from our naive scheduling algorithm, which as we saw closely approximates an optimal schedule, serves as a proxy for the upper bound.
We see that a speedup of 4.5x is not surpassed even including our disentanglement, even though we only look at theoretical bounds and do not impose limits on the degree of parallelism, e.g. a maximum number of threads.

\begin{figure}[t!]
\begin{subfigure}[t]{0.492\linewidth}
  \centering    
    \includegraphics[scale = 0.45]{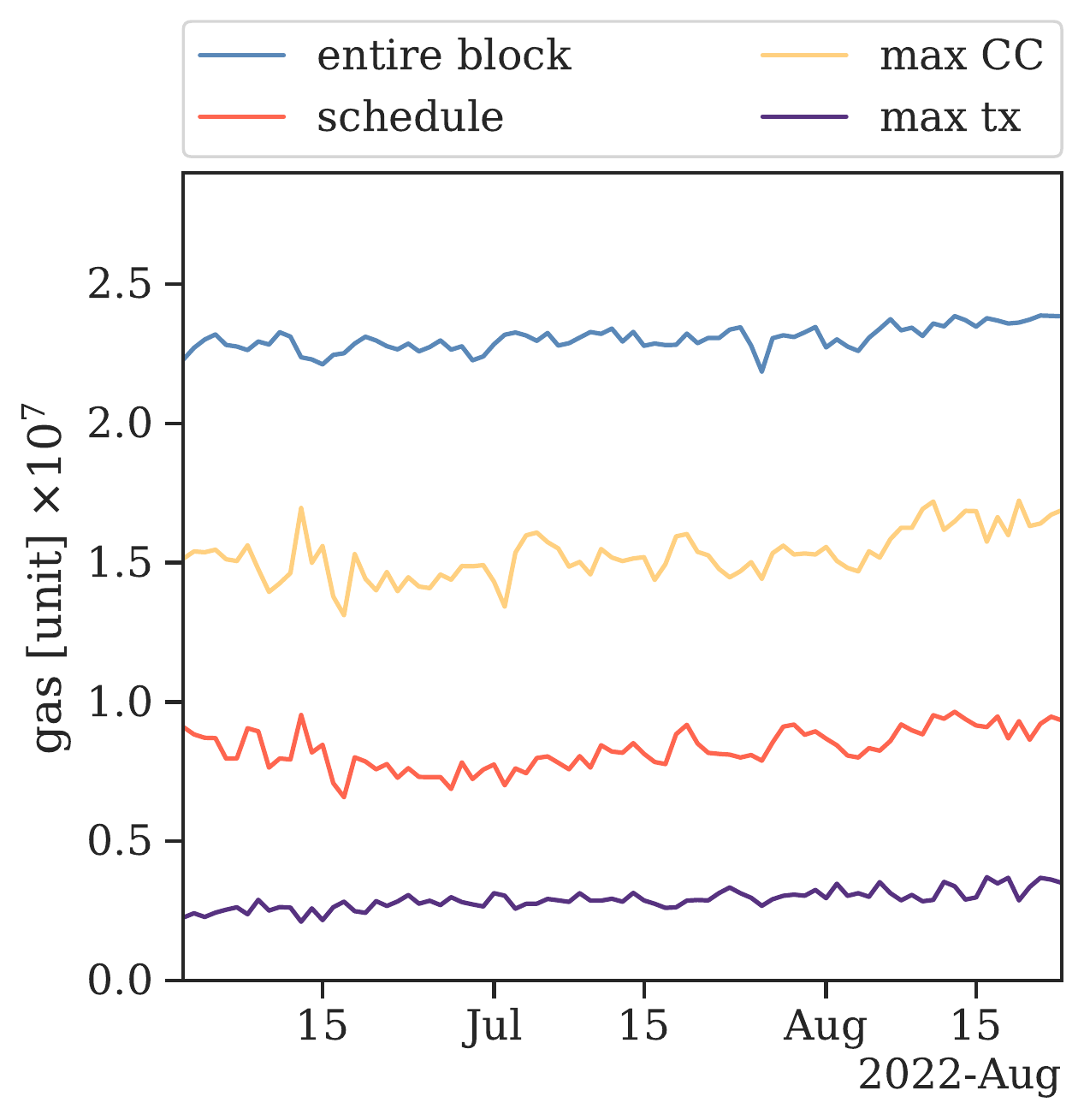}
    \caption{original transaction data} \label{fig:percentile90org}\vspace{0pt}
    (90th percentile)\vspace{2pt}
  \end{subfigure}%
  \hfill
  \begin{subfigure}[t]{0.492\linewidth}
  \centering    
    \includegraphics[scale = 0.45]{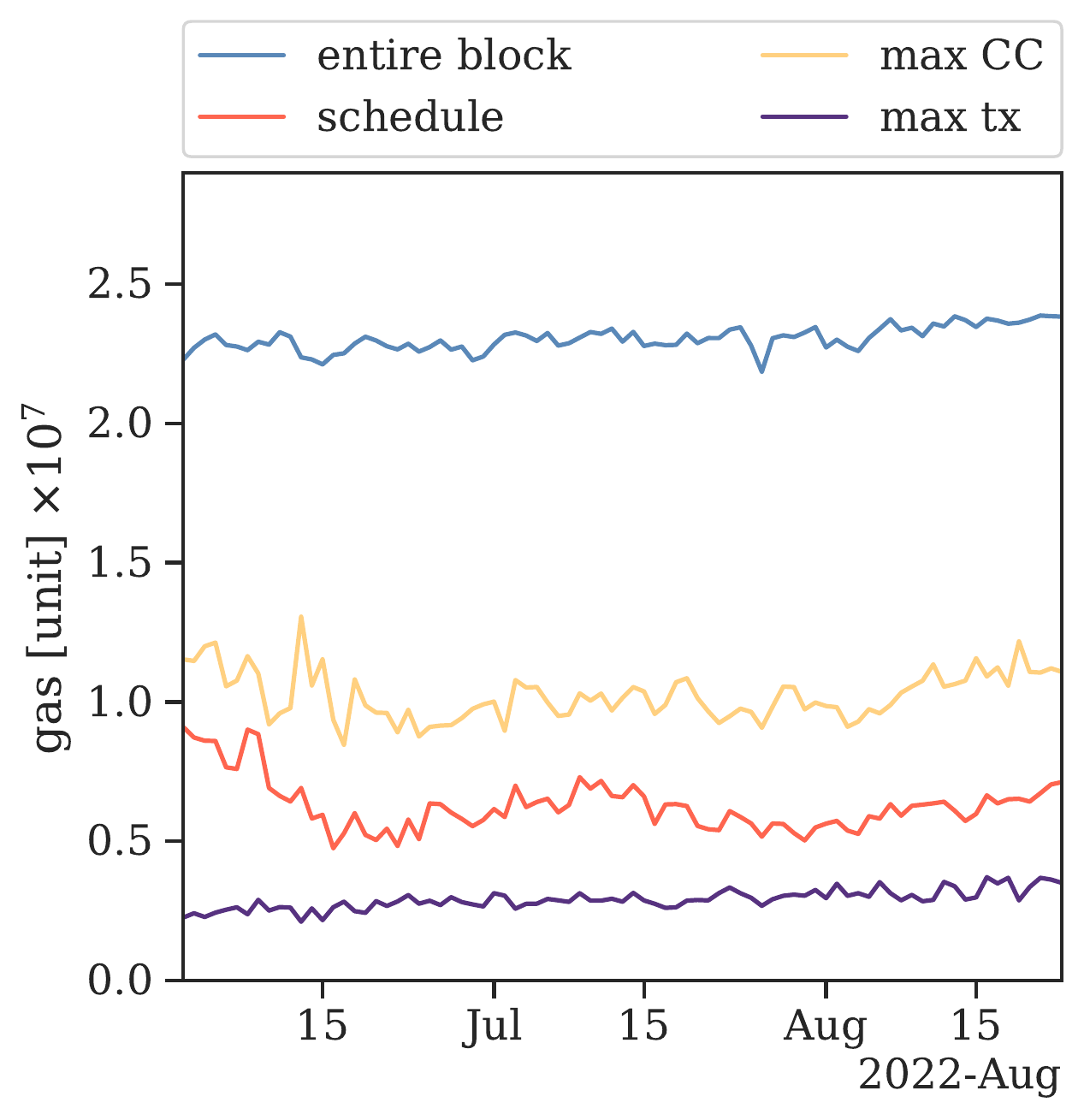}
    
    \caption{disentangled transaction data} \label{fig:percentile90disen}
    (90th percentile)\vspace{2pt}
\end{subfigure}
\begin{subfigure}[t]{0.492\linewidth}
  \centering    
    \includegraphics[scale = 0.45]{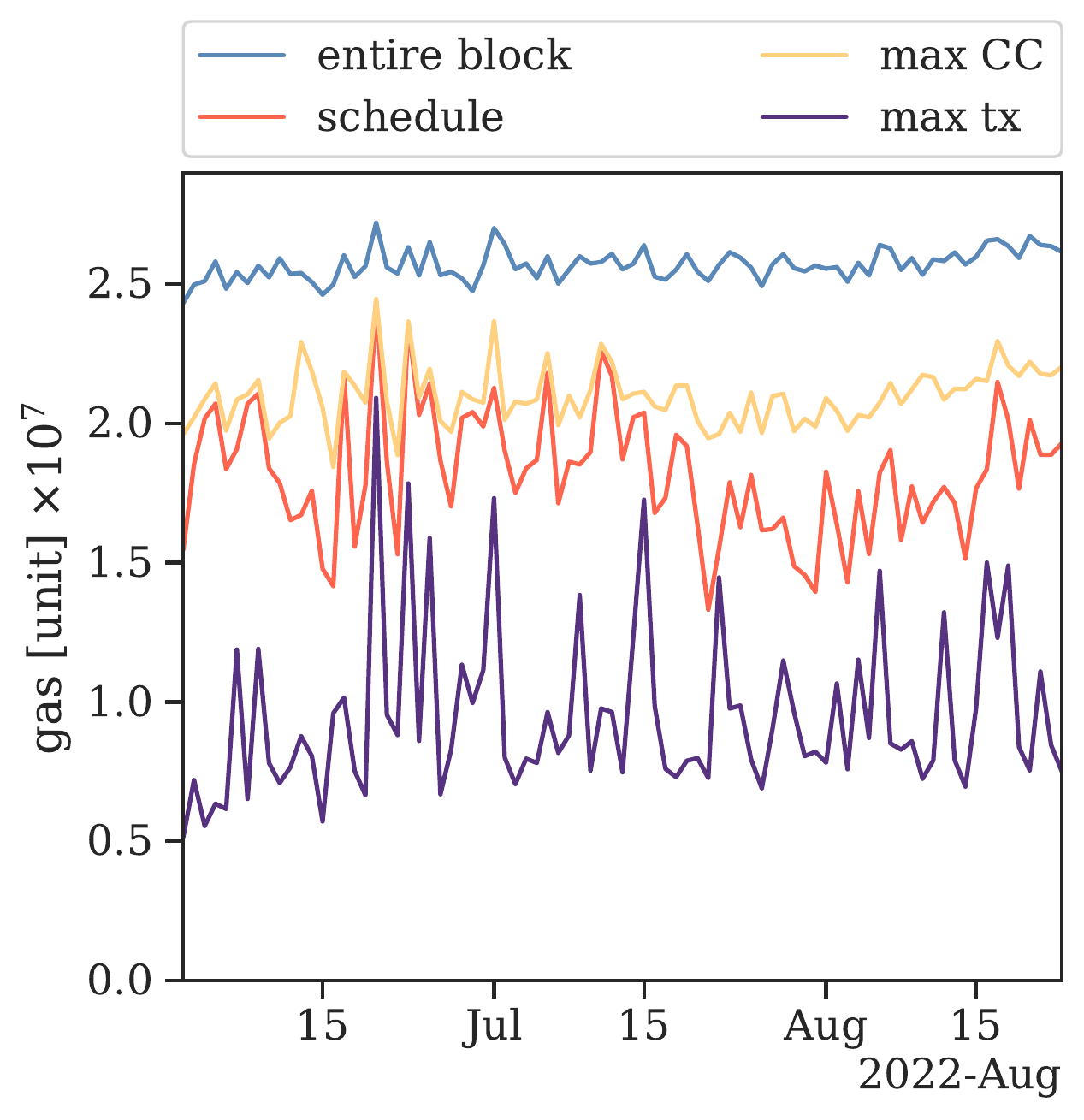}
    \caption{original transaction data} \label{fig:percentile99org}\vspace{0pt}
    (99th percentile)\vspace{2pt}
  \end{subfigure}%
  \hfill
  \begin{subfigure}[t]{0.492\linewidth}
  \centering    
    \includegraphics[scale = 0.45]{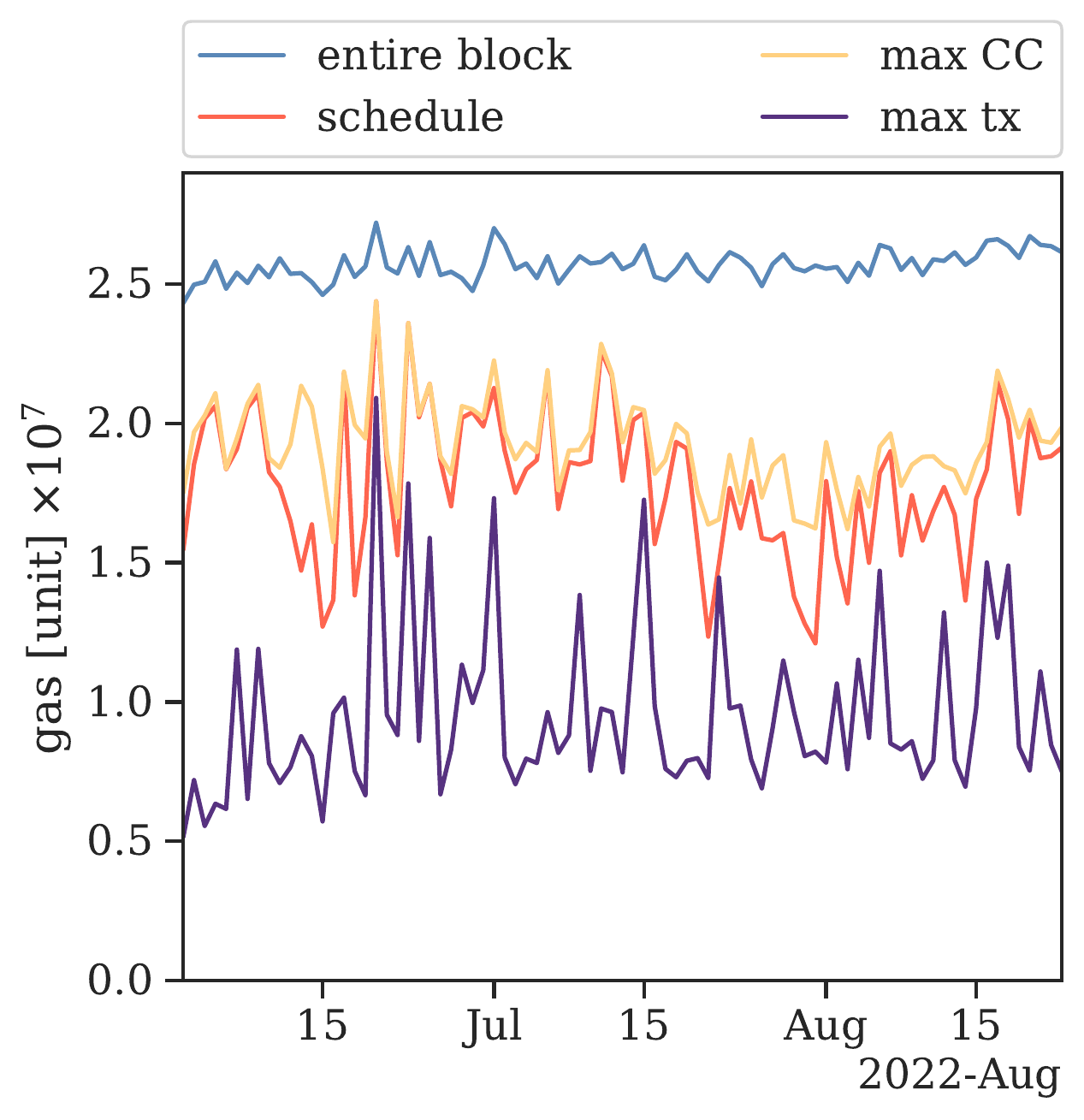}
    \caption{disentangled transaction data} \label{fig:percentile99disen}\vspace{0pt}
    (99th percentile)\vspace{2pt}
\end{subfigure}
\caption{We plot daily 90th and 99th percentile of the gas used by: (1) the entire block, (2) the block's heaviest connected component (CC), (3) our naive parallel schedule (sequential gas), and (4) the block's heaviest transaction. Figs.~\ref{fig:percentile90org} and~\ref{fig:percentile99org} analyze the original transaction data and Figs.~\ref{fig:percentile90disen} and~\ref{fig:percentile99disen} analyze the disentangled transaction data. Note that we use the sequential gas utilized by our schedule as a proxy for the size of the heaviest clique.}\label{fig:percentile}
\end{figure}

We plot the 90th percentile of the gas utilized for (1) the sequential execution of the entire block, (2) the heaviest connected component, (3) our naive schedule, and (4) the heaviest transaction, in Fig.~\ref{fig:percentile90org} for the original transaction data and in Fig.~\ref{fig:percentile90disen} for the disentangled transaction data.
Notice that in comparison to the daily mean (cf. Fig.~\ref{fig:recentgas}), the daily 90th percentile of gas used by all four measures is approximately a factor of two larger. Thus, 10\% of blocks are twice the size of the average block.
Note that EIP-1559 permits blocks to be at most twice the size of the target block size~\cite{2022eip1559}, and we are observing a significant proportion of blocks being built with a size very close to twice the target size.
We also observe many blocks that are almost empty at an equal rate and presume that this pattern of empty and overfull blocks emerged to manipulate the base fee introduced in EIP-1559.
The fact that we observe a similar increase in the 90th percentile of gas used by the heaviest connected component and our naive schedule, in comparison to the mean, indicates that in the 90th percentile of blocks, in terms of gas used, the level of connectedness is similar to that in an average block.
That is, the proportion of a block in the heaviest connected component and clique remains largely unchanged. Note that this is the case for both the original and disentangled transaction data.

If we turn to the 99th percentile of gas usage (cf. Figs.~\ref{fig:percentile99org} and~\ref{fig:percentile99disen}), a different picture paints itself.
We are no longer simply observing a proportional increase in gas usage across all four measures but instead observe widely different patterns.
Only the 99th percentile gas usage of the entire block appears stable, as it is limited by EIP-1559.
The 99th percentile of both the heaviest connected component and the heaviest clique is significantly larger than the average block size (cf. Fig.~\ref{fig:recentgasfig}).
This holds for both the original and the disentangled transaction data.
In the 99th percentile, our disentanglement has little impact on the heaviest cliques but still reduces the size of the largest connected components slightly.
We presume that this is due to the different nature of the transactions in the heaviest clique in these extraordinary cases.
Further, the 99th percentile of the largest transaction is as large as the mean block size (cf. Fig.~\ref{fig:recentgasfig}) and half the size of the 99th percentile of the block size.
Thus, the largest transactions make up significant proportions of a block and thereby hugely limit the parallelization potential of those blocks.

\section{Recent Ether Price Data}\label{app:ethprice}
\begin{figure}[h]
\begin{subfigure}[t]{0.492\linewidth}
  
  \centering    
    \includegraphics[scale = 0.45]{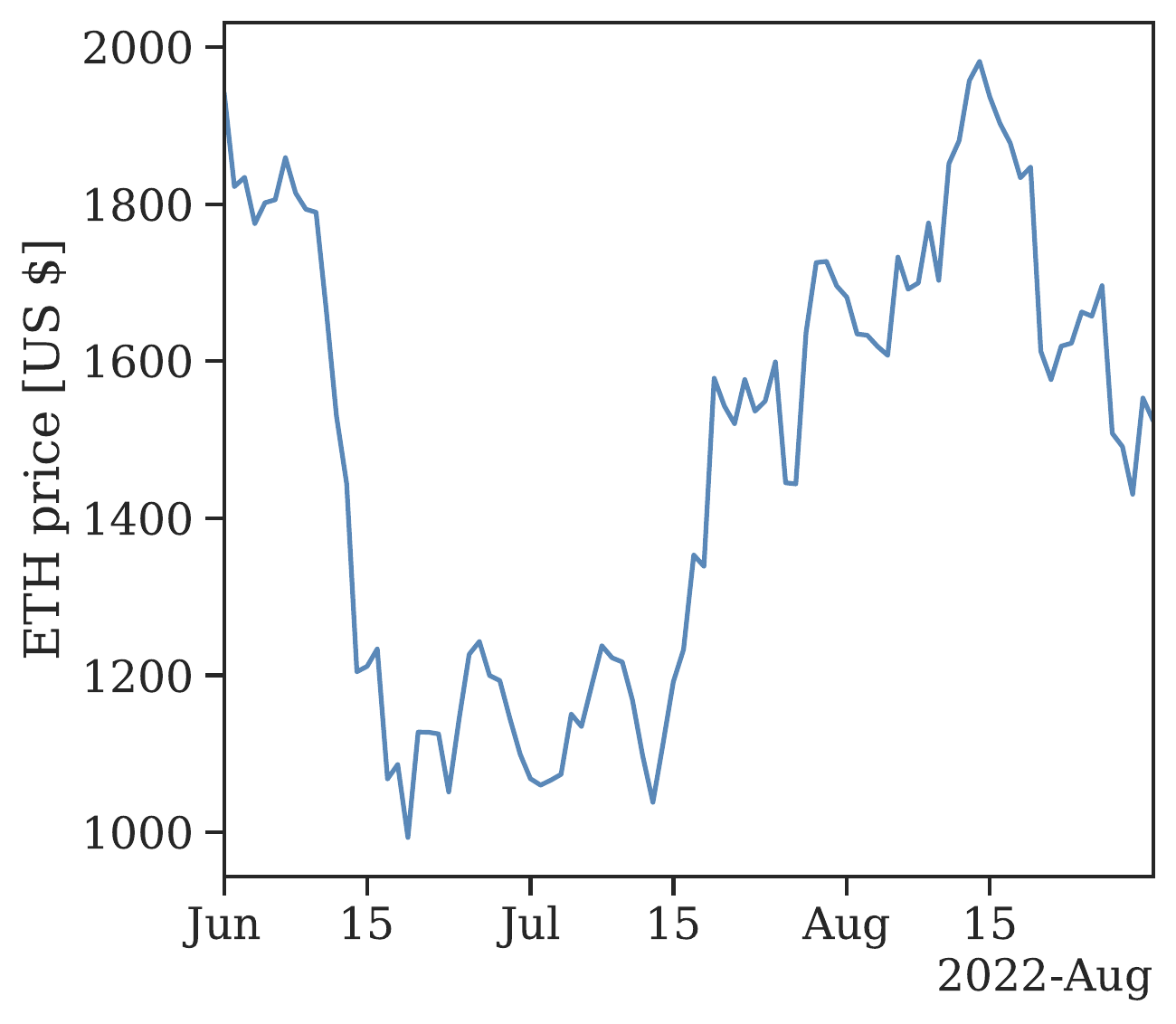}
    \caption{Ether price} \label{fig:ethprice}\vspace{0pt}
  \end{subfigure}%
  \hfill
  \begin{subfigure}[t]{0.492\linewidth}
  \centering    
    \includegraphics[scale = 0.45]{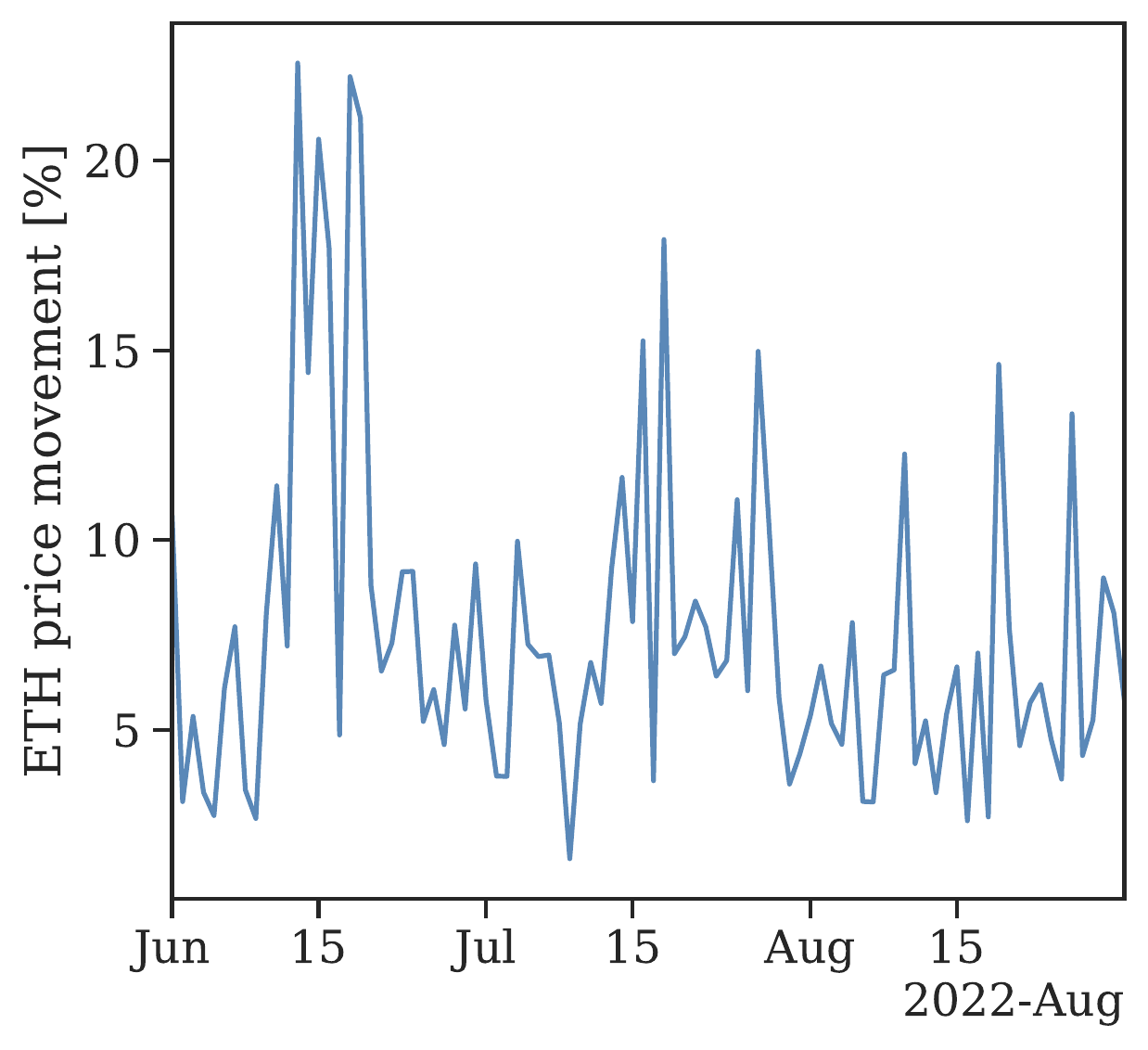}
    \caption{Ether price movements} \label{fig:ethmove}\vspace{0pt}
\end{subfigure}
\caption{We plot the Ether price and daily price movement from 1 June 2022 through 31 August 2022. The daily price movement compares the day's high to the day's low and is, therefore, a measure of price volatility.}\label{fig:eth}
\end{figure}

In Fig.~\ref{fig:eth}, we plot Ether's daily price and daily price movement from 1 June 2022 to 31 August 2022.
Fig.~\ref{fig:ethprice} visualizes the Ether's daily open price ($p_{{open}}$) and daily close price ($p_{{close}}$).
The daily price movement, which we plot in Fig.~\ref{fig:ethmove}, is a measure of the daily price volatility and is given by
$$ \frac{p_{{high}}- p_{{low}}}{p_{{low}}},$$
where $p_{high}$ is the day's highest Ether price and $p_{low}$ is the day's lowest Ether price~\cite{Berg2022empirical}.

\end{document}